\documentclass{article}
\usepackage[english]{babel}
\usepackage[latin1]{inputenc}
\usepackage{graphicx}
\usepackage{amssymb,amsmath}
\usepackage{pstricks}
\usepackage{dcolumn}
\usepackage{bm}
\usepackage{geometry}
\title{Formation of spanwise vorticity in oblique turbulent bands of transitional plane Couette flow, part 2: modelling and stability analysis.}
\date{\today}
\author{Joran Rolland\footnote{rolland@iau.uni-frankfurt.de}
\footnote{LadHyX, UMR 7646 CNRS, Palaiseau 91128 France, \emph{current address:} IAU, Frankfurt Goethe University, Altenh\"oferallee 1, 60438 Frankfurt am Main, Deutschland.}}

\begin{document}
\maketitle


\begin{abstract}
This article presents a modelling of the formation of spanwise vorticity in the turbulent streaks of the oblique bands and spots of transitional plane Couette flow. A functional model is designed to mimic the coherent flow in the streaks. The control parameters of the model are extracted from Direct Numerical Simulations (DNS) statistical data. A Reynolds stress is proposed to study the effect on the instability of this additional force maintaining the baseflow. Local (quasi-parallel) temporal stability analysis is performed on that model to investigate the linear development of the spanwise vorticity. Results show that average profiles, even if they have an inflection, are stable: the shear layers inside the velocity streaks are responsible for the vorticity formation. Emphasis is put on the convective or absolute nature of the instability, depending on the location in the band. This shows that a transition from a convective to an absolute instability occurs in the zone in between fully turbulent and laminar flow. The group velocity of the most unstable modes compare well to the advection velocity of spanwise vorticity measured in DNS. This investigation is completed by the global (non-parallel) stability analysis of a typical band case. Eventually, the possible cycles of sustainment of localised low Reynolds number turbulence of shear flows are discussed in the light of these results.
\end{abstract}
\begin{flushleft}
 transition, Wall-bounded turbulence, shear flow instabilities
\end{flushleft}
\begin{flushleft}
47.27.Cn, 47.27.N-, 47.20.Ft
\end{flushleft}
\section{Introduction}

\begin{figure}
\centerline{{\large \textbf{(a)}}\includegraphics[height=3cm]{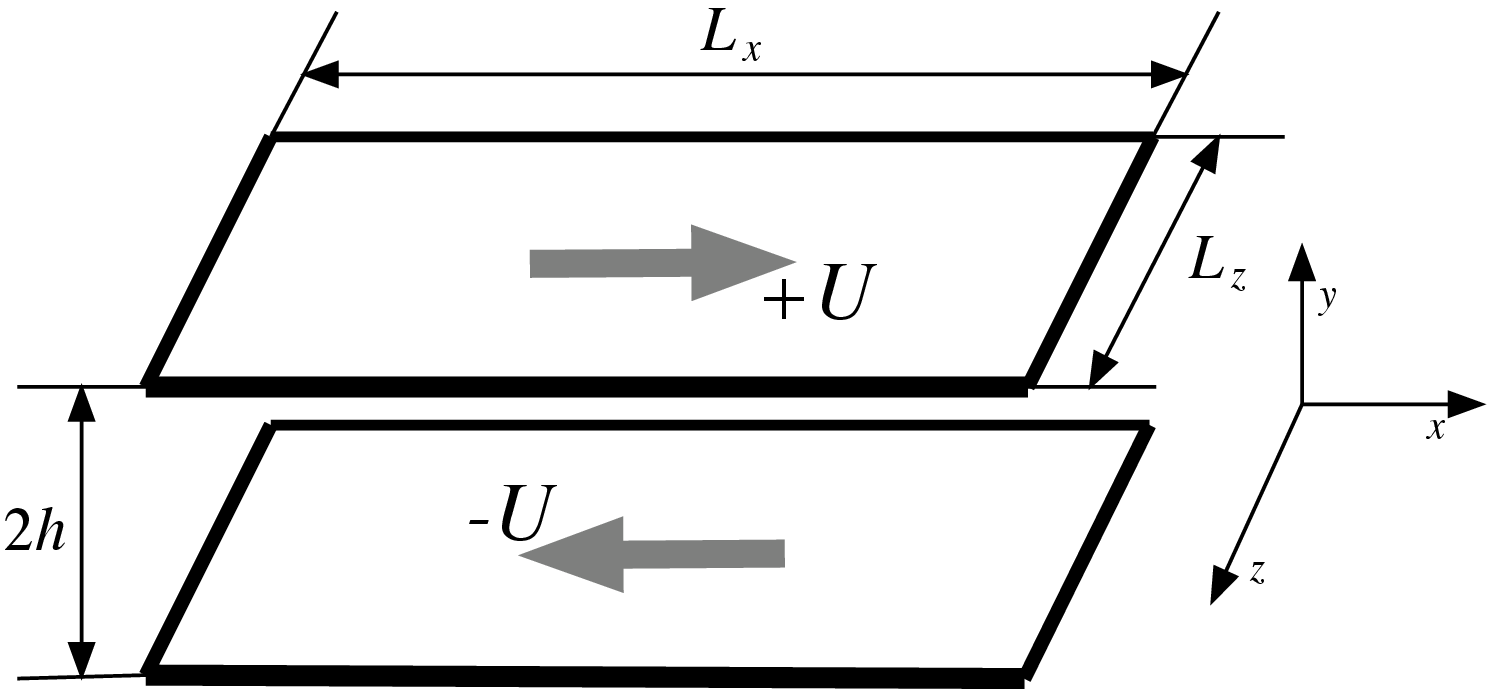}\vspace{1mm}
{\large \textbf{(b)}\hspace{1mm}\includegraphics[width=3.8cm,clip]{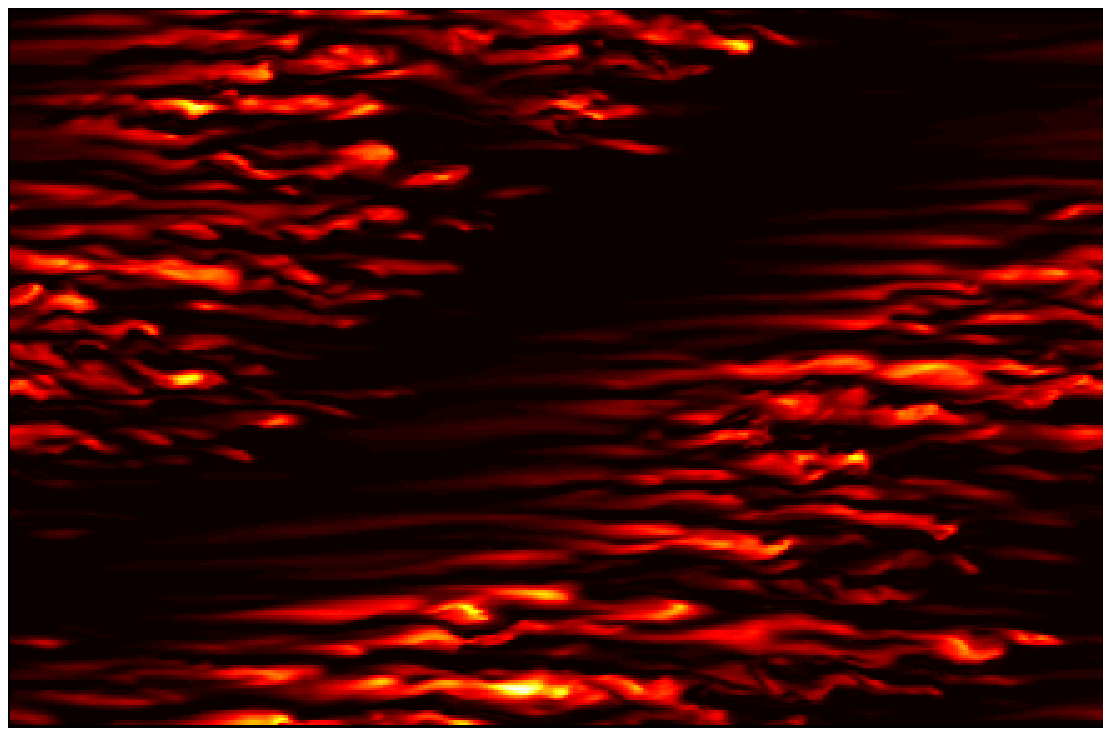}\hspace{1mm}\textbf{(c)}\hspace{1mm}}\includegraphics[width=3.8cm,clip]{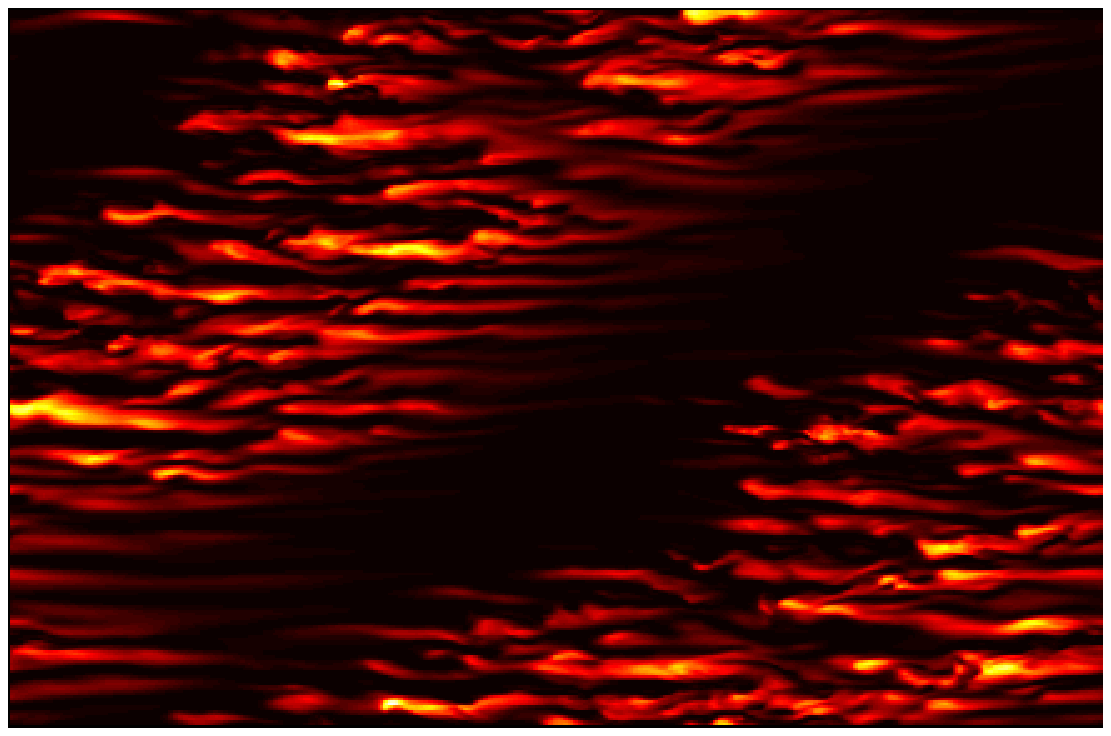}}
\caption{(a): Sketch of plane Couette flow. Example of turbulent oblique band, colour plot of $\bf{v}^2$ in a $y=-0.62$
plane (b) and $y=0.62$ plane (c) in a periodical domain ($L_x=110$, $L_z=72$, $R=370$). }
\label{f1}\end{figure}

  In its route toward turbulence, plane Couette flow (PCF), the flow between two parallel moving planes (Fig.~\ref{f1} (a)), displays laminar-turbulent coexistence. This coexistence takes the form of stationary oblique bands in the range of Reynolds number $R\in[325;415]$ (Fig.~\ref{f1} (b,c)), where $R=hU/\nu$, with $h$ the half gap, $U$ the velocity of the planes, and $\nu$ the kinematic viscosity. Under The Reynolds number $R_{\rm g}=325$, turbulence is not globally sustained. Above the Reynolds number $R_{\rm t}=415$, turbulence invades the whole domain. The bands correspond to a sinusoidal modulation of low Reynolds number turbulence \cite{prigent,BT07,RM,BT11}. In a larger range of Reynolds number, the growing or meta-stable turbulent spots are a time-dependent form of oblique laminar-turbulent coexistence \cite{lJ}.  Unlike the self-sustaining of turbulence in fully turbulent zones, the mechanisms responsible for the cohabitation of laminar and turbulent flow are still under investigation by mean of modelling \cite{BPPF,B} or DNS \cite{SK,DWK,isp1}. It is all the more intriguing that one can see turbulence entirely collapse at $R\lesssim 415$ in small domains \cite{DLS}. Meanwhile turbulence is able to sustain itself at smaller and smaller $R$ as the size $L_x\times L_z$ of the domain is increased \cite{JP}. The value $R_{\rm g}$ is reached only when the system is large enough so that oblique bands manifest themselves \cite{shi,JP}.

  DNS of Hagen--Poiseuille Pipe Flow (PPF) \cite{SK,DWK} and PCF \cite{isp1,ispspot}
recently brought a streamwise shear instability into attention. The
phenomenon is reported in plane Poiseuille flow as well \cite{ATK}.
It was shown \emph{via} DNS of PPF that the secondary instability of the streaks
occurred in the trailing edge of puff and slugs. This secondary instability led to azimuthal vorticity. The vortices are advected upstream, toward
developed turbulence in the case of low Reynolds number metastable puffs ; or downstream
toward the laminar region, in the case of high Reynolds number expending slugs. A lagrangian method was used to compute the speed of coherent structures in laminar-turbulent interfaces of puffs in pipe flow by Holzner \emph{et al.} \cite{lag}. This confirmed the less technical earlier measurements \cite{SK,DWK}. Shimizu \&
 Kida argued that the developed instability fed back on the turbulent puff, which in turn regenerate the velocity streaks, thus creating a cycle of sustainment of the puff \cite{SK}. Duguet \emph{et
al.}  studied the phenomenon at several Reynolds numbers
and related the velocity of vortices relatively to the turbulent
region to the growth and recess of turbulence region \cite{DWK}. In the case of PCF, part $1$ showed that the wall normal shear layers found inside the streaks developed into rolls, identified by their spanwise vorticity $\omega_z$ \cite{isp1}. Besides, it showed that said rolls are advected either along the band at the velocity of the wall normal averaged large scale flow or toward the laminar zone where they are dissipated.

  In order to understand the formation of vorticity, Shimizu \& Kida \cite{SK} and Duguet \emph{et
al.} \cite{DWK} referred to the local instability of a two dimensional shear layer \cite{V,BS}. Indeed, the assumption of a frozen shear layer or frozen corrugation undergoing an instability has been very efficient in the former studies of breakdown of velocity streaks \cite{W,schhu,KJUP}. However they made no systematic study of the advection velocity of the perturbations. A parametric study, as a function of the position in the modulated laminar-turbulent coexistence, is necessary in order to model the modulated advection velocity measured in DNS \cite{isp1}. A specific study is further justified by the fact that PCF has a spanwise extension and that the vorticity can be advected in both spanwise and streamwise directions.

  In order to understand the development of the shear layers of the streaks into spanwise vorticity and its advection, we turn to linear stability analysis. This type of approach is justified since numerous studies showed that the velocity streaks remained coherent over long periods of time and that their evolution was well described by linear analysis \cite{W,schhu,KJUP,isp1}. Moreover, the scale separation in PCF between the wavelength of bands \cite{isp1} and spots \cite{PRL,ispspot} ($\simeq 50$ gaps) and the typical length scale of the perturbation to the velocity streaks (a few gaps), calls for a quasi-parallel (or local) study of the instability \cite{HM,DR}. The computation of group velocities is central to that framework. It determines the convective or absolute nature of the instability, and therefore if and where the perturbations are advected. The non-parallel (or global) stability analysis, which considers the whole band, is the final stage of such approaches.

  One needs to determine the baseflow for such an analysis. This raises two questions because this baseflow is not the simple solution of a steady Navier--Stokes problem. Firstly, one must extract the characteristics of the baseflow from DNS. This is typically done by using a functional wall normal dependance deriving from average data, and determine the distribution of control parameters from DNS data \cite{metal,schhu}. Said parameters yield a baseflow from the profiles averaged in time and along the diagonal direction of the band. Indeed, these parameters take into account the spanwise corrugation and the turbulent fluctuations. Secondly, one may have to take into account a Reynolds stress force maintaining the velocity streaks \cite{metal} as well as the slow spatial evolution of the baseflow \cite{B_w}. In that matter, the study of the convective to absolute transition in the wake of a cylinder and of the self oscillating structure that ensued can enlighten us. Such a transition can occur in a local study of the wall normal shear layer of velocity streaks. In both cases, one has to sample the velocity profile. The studies of the wake flow showed that the local and the global analyses, without stress and using a baseflow sampled from DNS, reproduced numerical and experimental results very well \cite{P,B_w,SL}. However, some cases are not as simple \cite{SL}. It is not unreasonable to construct our baseflow from DNS samples, however, the effect of such a force must be tested before including or neglecting it in a systematical study.

  In order to investigate the secondary instability, the article is organised as follow: The first section (\S~\ref{model}) is concerned with the description of the model of the turbulent baseflow (\S~\ref{modx},~\ref{bf}), based on sampling of DNS data (\S~\ref{param}). The section also describes a model of Reynolds stress (\S~\ref{mrs}). The second section (\S~\ref{lsp}) describes the framework of linear stability analysis used here (\S~\ref{fw}) and our numerical procedure (\S~\ref{num}). The results of the analysis are presented in the next section (\S~\ref{res}). Local one component analysis is presented first (\S~\ref{res1}), then local two component analysis (\S~\ref{res2}), and eventually the global analysis (\S~\ref{glob}). In the conclusion (\S~\ref{disc}), the results are summed up and put back in the context of the turbulent-laminar cohabitation, and the possibility of self-sustaining processes of laminar-turbulent cohabitation is discussed.

\section{modelling \label{model}}

\subsection{Notations}
\begin{figure*}
\centerline{\includegraphics[width=12cm,clip]{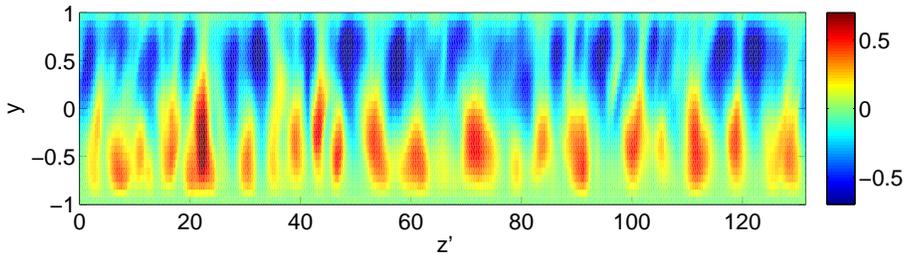}}
\caption{Colour levels of the streamwise velocity $v$ (the laminar flow is subtracted), along a diagonal, averaged over a period of time $T=25$
(inside the turbulent zone).}
\label{figalpha}
\end{figure*}
  Plane Couette flow is the flow between two parallel moving plates, moving at speed $\pm U \mathbf{e}_x$ at positions $y=\pm h$ (Fig.~\ref{f1} (a)). We used a dimensionless version of the problem in order to simplify comparisons between configurations. The lengths are made dimensionless by $h$, the velocities by $U$ and the time by $h/U$. The spanwise direction is termed $z$. The oblique bands (Fig.~\ref{f1} (b,c)) lead to the introduction of an additional diagonal direction $z'$ parallel to the band. Note that the properties of the velocity streaks are statistically invariant along that direction \cite{BT07}: in $y,z'$ planes, one finds only pseudo-laminar flow, turbulent flow (Fig.~\ref{figalpha}) or intermediate flow \cite{isp1}(sometimes called overhanging \cite{cole66}).

  Using the kinematic viscosity $\nu$, the Reynolds number of the flow is $R=hU/\nu$. Together with the sizes $L_x$ and $L_z$ (Fig.~\ref{f1} (a)) it controls the state of the turbulent flow. Note that in all our DNS, the in plane boundary conditions are periodic: $L_x$ and $L_z$ are therefore the wavelength of the band. (Fig.~\ref{f1} (b,c))

\subsection{Long wavelength streamwise flow\label{modx}}

 We first present a description of the modulation of turbulence focusing on the intermediate (or overhanging  \cite{cole66}) zone between laminar and turbulent flow. In that part of the flow, turbulent flow is found for $y>0$ (resp. $y<0$) while laminar flow is found for $y<0$ (resp. $y>0$). Describing the large-scale flow is fundamental in order to determine the type of profile to be used in the stability analysis and the streamwise dependence of the global flow. This first description is averaged in the $z'$ direction and does not include the small spanwise length-scale ($\simeq 2h$) of the velocity streaks as well as the turbulent fluctuations.
We use data sampled in DNS of plane Couette flow in a domain of size $L_x=110$, $L_z=72$, at Reynolds number $R=370$, in the middle of the existence range of the bands. The DNS is performed using the {\sc Channeflow} code by J. Gibson \cite{gibs} with a resolution sufficient to give reliable quantitative results ($N_{x,z}/L_{x,z}=4$, $N_y=27$) \cite{MR,dsc10}. The procedure to obtain statistically steady bands is the same as in part $1$: wall turbulence is obtained at $R=500$, the Reynolds number is decreased to $R=370$, and the flow is allowed to evolve long enough for the band to stabilise \cite{isp1}.

\begin{figure}
\centerline{\includegraphics[height=5cm]{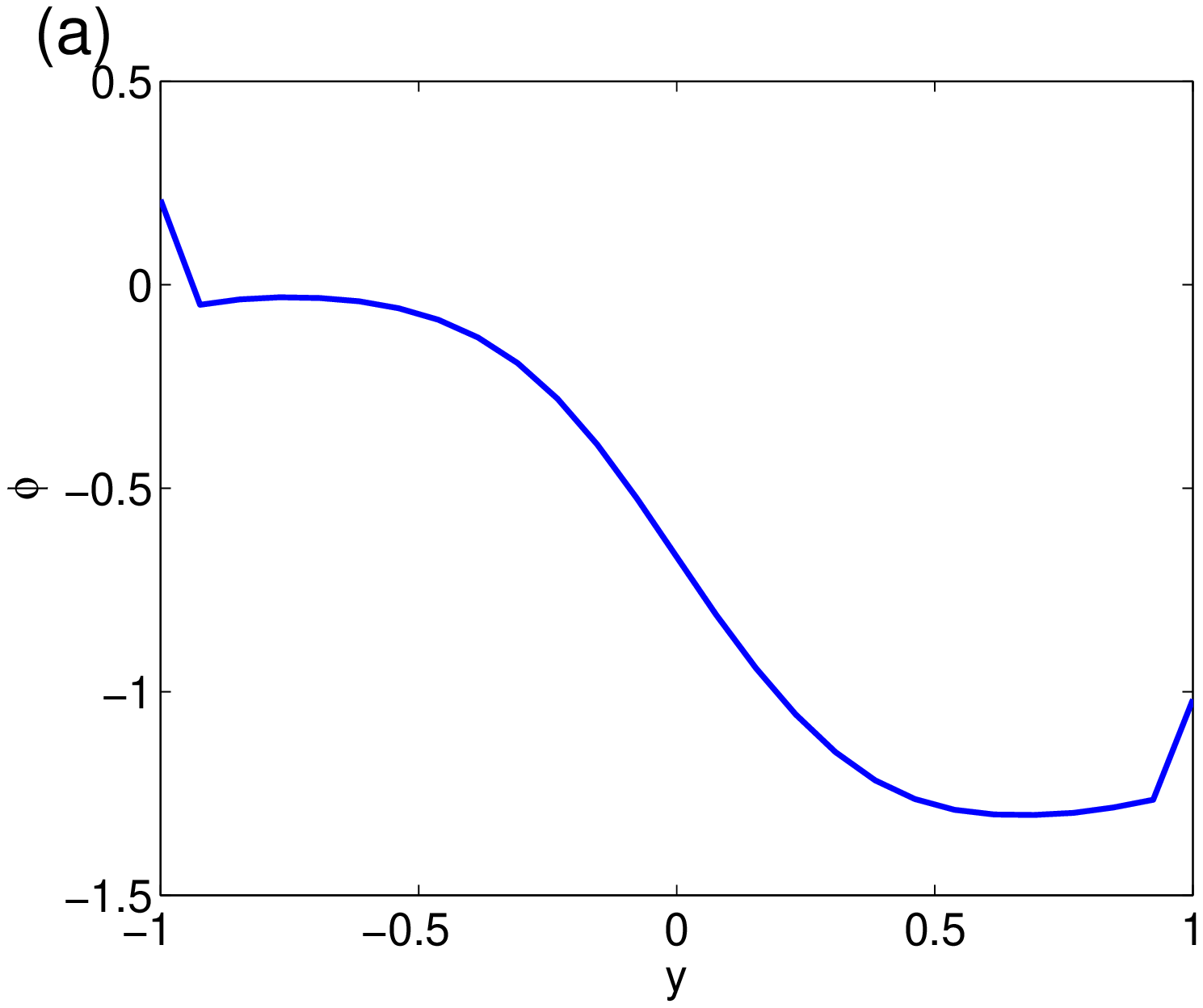}\includegraphics[height=5cm]{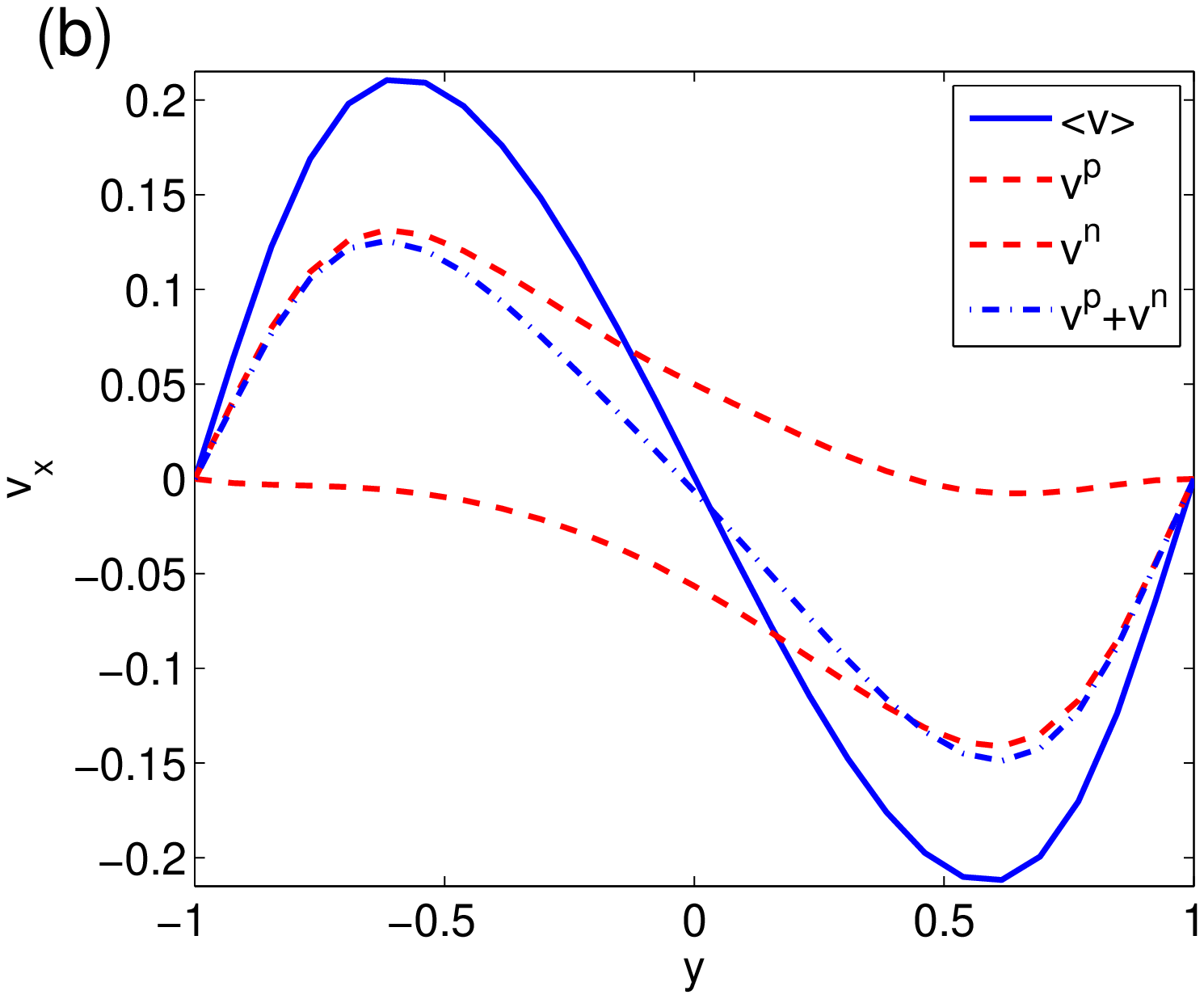}\includegraphics[height=5cm]{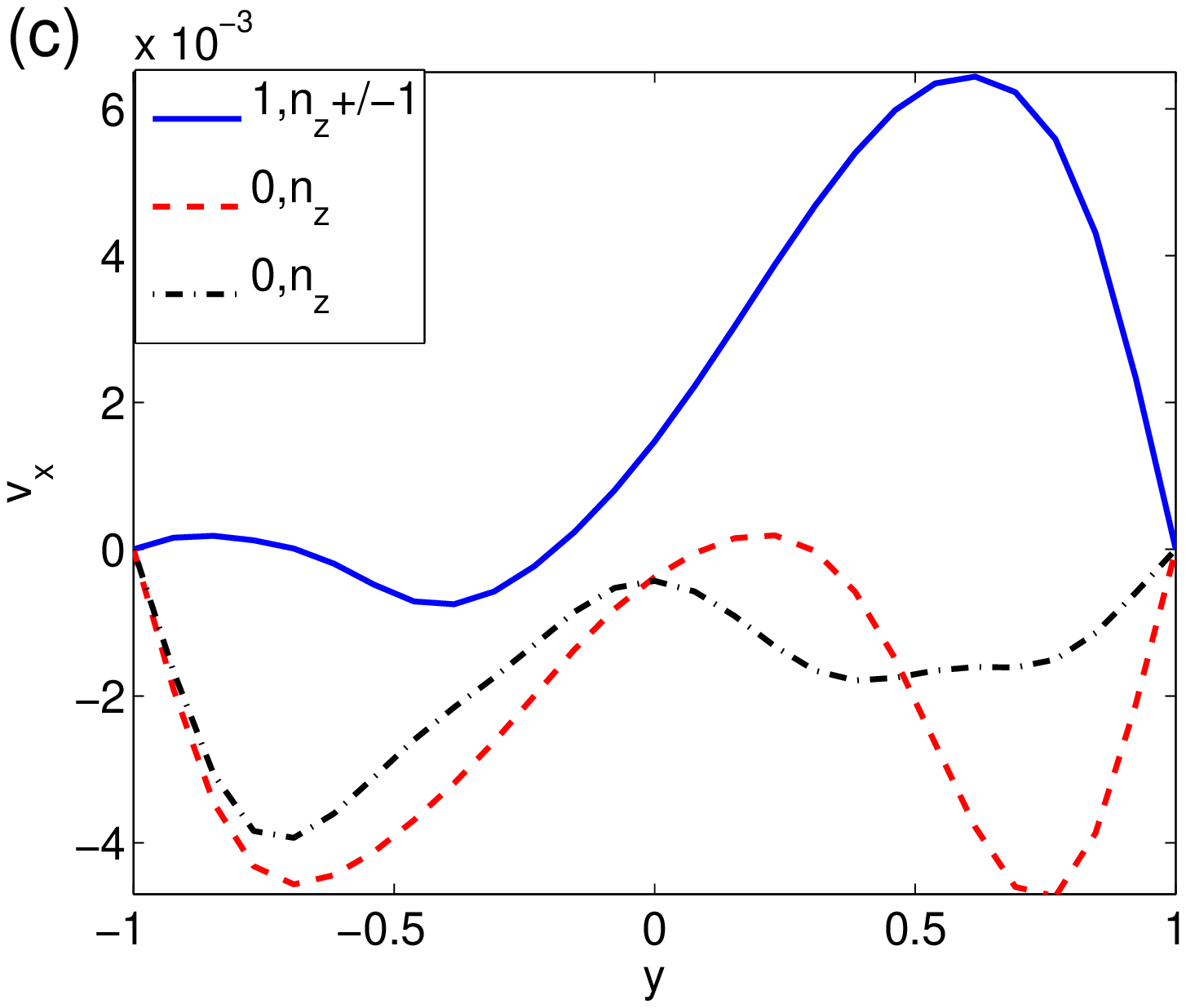}}
\caption{(a): Phase of the band mode as a function of $y$ $k_x=1$, $k_z=1$.(b): Profiles extracted from the band mode. (c) Profiles corresponding to the spanwise short wavelength dependence.}
\label{mdl}
\end{figure}

  A Fourier transform of the streamwise velocity field $v_x$ is performed at each wall normal position $y$. The results are very similar to the profiles computed by Barkley \& Tuckerman in a tilted domain \cite{BT07}. The modes of wavenumbers $(k_x,k_z)=(0,0)$ and $(1,\pm 1)$ describe the large scale modulation. In our case, only the modes $0$ and $1,1$ exist (Fig.~\ref{f1} (b,c)). The fundamental mode $(1,1)$ is denoted $\hat{v}(y)\equiv \widehat{v_x}(k_x=1,y,k_z=1)$. They wrote the large scale modulation $\tilde{V}_x$
\begin{equation}
\tilde{V}_x=g_0(y)+g_c(y)\cos\left(2\pi \left(\frac{x}{L_x}+\frac{z}{L_z}\right)+\varpi\right)+g_s(y)\sin\left(2\pi \left(\frac{x}{L_x}+\frac{z}{L_z}\right)+\varpi\right)\,,\label{eqBT}
\end{equation}
with $\varpi$ a phase. We rewrite this velocity field so that the functions of $y$ correspond to $V_x$ in the intermediate zones. We first use the complex formulation of the Fourier transform.  The phase $\phi (y)\equiv \arg(\hat{v}(y))$ of the mode $1,1$ is displayed in figure~\ref{mdl} (a). One typically has $\phi(y)=\phi_1$ if $y<0$
and $\phi(y)=\phi_2$ if $y>0$. The phase $\phi_1$ at $y=-1$ is used as a reference and set to
zero. The phase difference $\psi=\phi_2-\phi_1$ is extracted. It yields the spatial shift of all quantities between the $y<0$ and $y>0$ part of the flow, in particular the one leading to the intermediate regions. Focusing on the two intermediate regions, this is rewritten using $\psi$
\begin{equation}\tilde{V}_x=y+\langle v\rangle+v^{\rm p}\cos\left(2\pi \left(\frac{x}{L_x}+\frac{z}{L_z}\right)\right)+v^{\rm n}\cos\left(2\pi \left(\frac{x}{L_x}+\frac{z}{L_z}\right)+\psi\right) \,,\end{equation}
with $\langle v\rangle$ the spatial average (or mode $0,0$) and $v^{{\rm p}, {\rm n}}$ which describe the typical profile in the intermediate regions: $v^{\rm p}>0$ for $y<0$, $v^{\rm p}=0$ for $y>0$ and $v^{\rm n}<0$ for $y>0$ and $v^{\rm n}=0$ for $y<0$. Beware that there is a global position shift with respect of description Eq.~(\ref{eqBT}) ($\varpi\ne0$).
The relation between the complex $1,1$ Fourier mode $\hat{v}$ and this description is
\begin{equation}\hat{v}=\frac{\bar{v}^{\rm p}+\cos(\psi)\bar{v}^{\rm n}+\imath\sin(\psi) \bar{v}^{\rm n}}{2}\,. \end{equation}

The two profiles $\bar{v}^{{\rm p},{\rm n}}$ are then computed by inverting this equation.
They are displayed in figure~\ref{mdl} (b).
The profiles verify $\langle v\rangle \propto \bar{v}^{\rm p}+\bar{v}^{\rm n}$ (figure~\ref{mdl} (b)) and the centro-symmetry $v^{\rm p}(y)=-v^{\rm n}(-y)$ \cite{BT07,ispspot}. The long wavelength dependence can therefore be rewritten
\begin{equation}\tilde{V}_x= y+ v^{\rm p}\left(1+\cos\left(2\pi \left(\frac{x}{L_x}+\frac{z}{L_z}\right) \right)\right)+v^{\rm n}\left(1+\cos\left(2\pi \left(\frac{x}{L_x}+\frac{z}{L_z}\right) +\psi\right)\right)\,.\label{eqmod2}\end{equation}
Numerical data show that $\psi\simeq \pi/2$. Meanwhile, the global shift with respect to the description of Eq.~(\ref{eqBT}) is approximately $\varpi\simeq-\pi/4$. This procedure extracts the average flow in the intermediate zone,
similarly to the conditional average performed in part 1 \cite{isp1}.
This emphasises the fact that the average large scale flow can be written with a linear combination of only one wall normal profile, symmetry operations and a modulation of amplitude. This fact has been pointed out for the average velocity profiles in spots \cite{ispspot}. In fact, higher wave number modes (for instance $k_x=0,1$, and large $k_z$ modes, Fig.~\ref{mdl} (c)) can also be described by a linear combination of such functions. These modes are typically those involved in the spanwise modulation of the velocity streaks at wavelength $\lambda_z =O(h)$.

\subsection{functional model\label{bf}}

  We construct a baseflow which mimics the profiles $\bar{v}^{{\rm p},{\rm n}}$. Former studies showed several important
parameters of the velocity streaks: amplitude, shear layer thickness and position of the maximum  \cite{W,schhu,metal}. We choose the formulation
 \begin{equation} v^{\rm n}\rightarrow h_1^{d}(y)\equiv-\Big(-\tanh(d(-1-s_1)) +\tanh(d(y-s_1))\Big)(1-\exp(-0.05(1-y)))\,,\label{profa1}\end{equation}
\begin{equation} v^{\rm p}\rightarrow h_2^{d'}\equiv\Big(-\tanh(-(1-s_2)d')  +\tanh(-d'(y-s_2))\Big)(1-\exp(-0.05(1+y)))\,.\label{profa2}\end{equation}
  An hyperbolic tangent dependence is chosen to represent the shear layer around $y=0$.
The parameters $d$ and $d'$  represent the inverse of the shear
layer thickness at $y=0$. They are the two control parameters of these
profiles. Note that these functions are not derived from first principles. Instead, they are an educated guess. They will be used in a normalised form, so that the prefactor that multiplies them directly gives the value of their maximum. Using this simple formulation, the profiles $v^{\rm p,\rm n}$ can be fitted with very few parameters whose meaning is clear.

\begin{figure}
\centerline{\includegraphics[height=5cm]{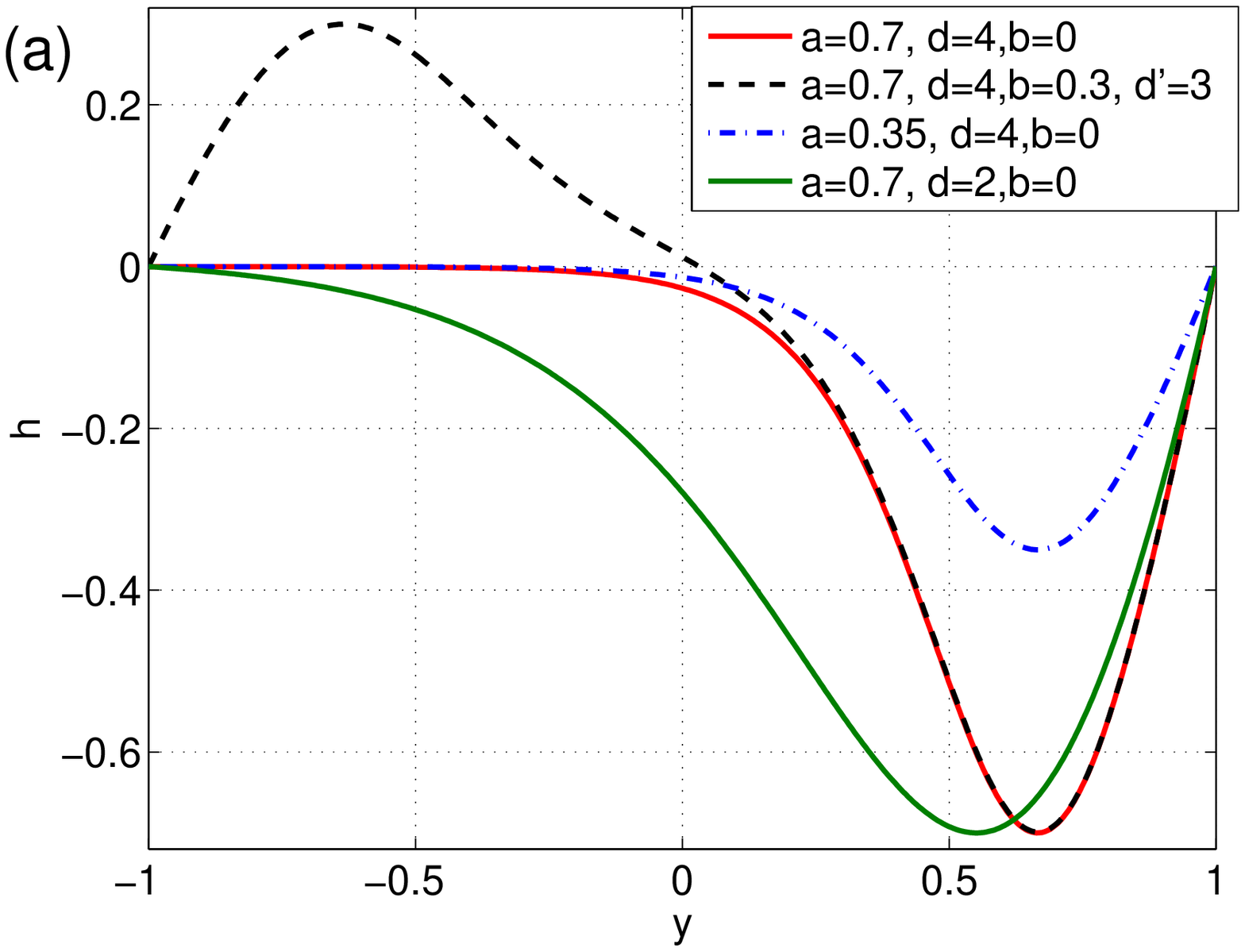}\includegraphics[height=5cm]{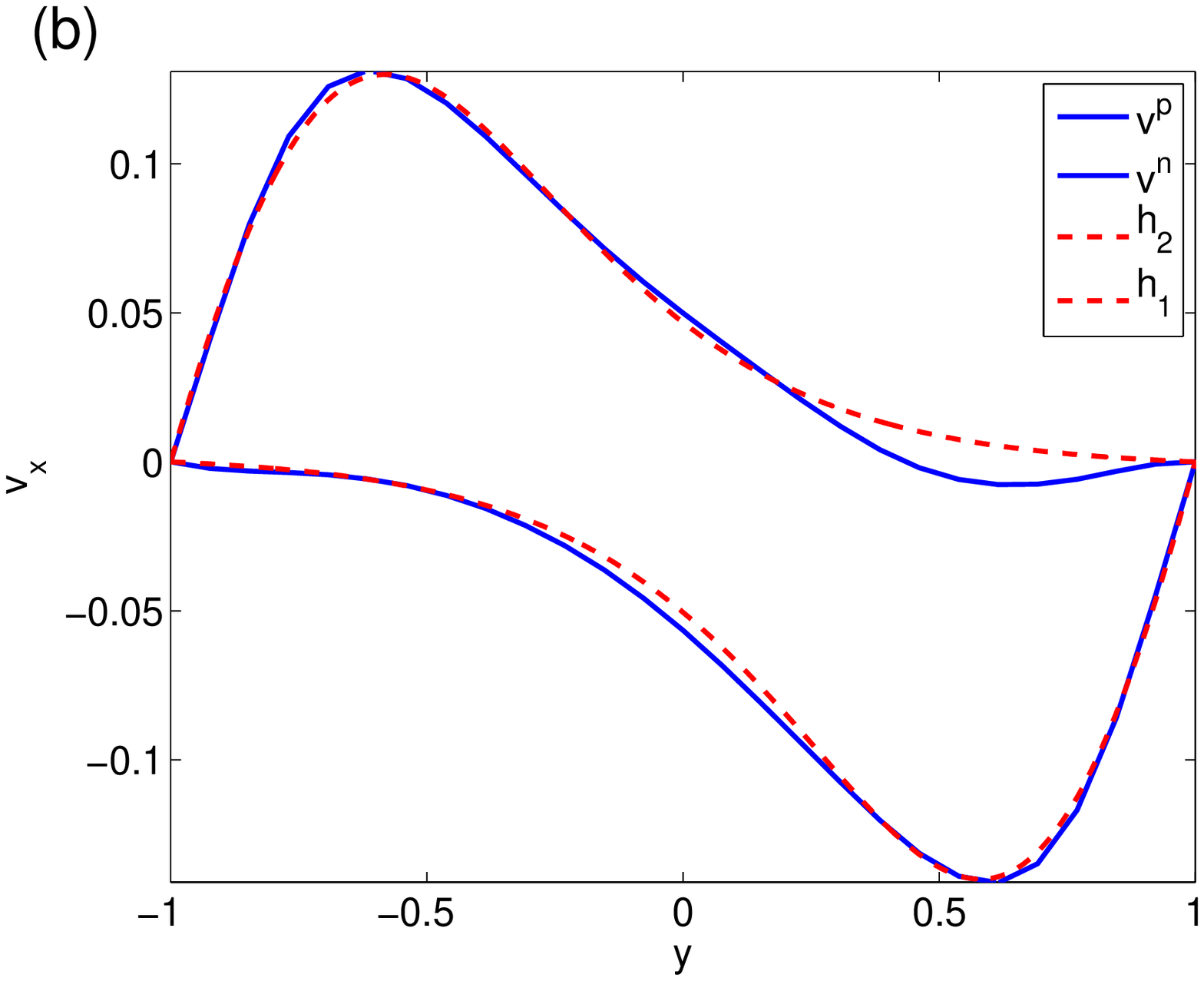}
\includegraphics[height=5cm]{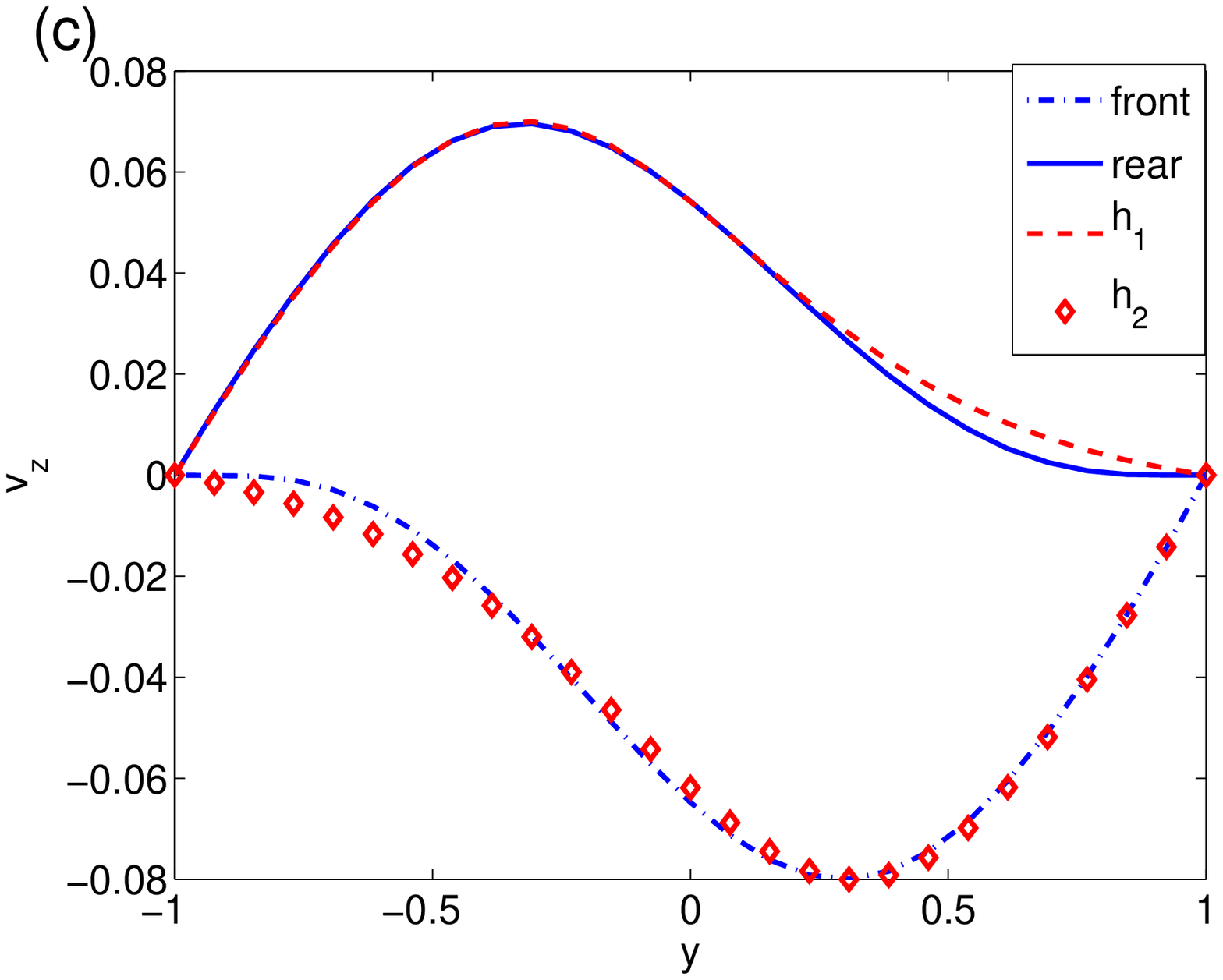}}
\caption{(a): examples of linear combination of functions $ah_1^d/\max_y(|h_1^d|)+bh_2^{d'}/\max_y(h_2^{d'})$ for four sets of values $a,d,d,d'$, demonstrating the presence or lack of a backflow, the changes in amplitude $a$ and the changes in inverse shear layer thickness $d$.(b) Examples of base flow velocity $v_x$ for parameters,
compared to $v^{{\rm p},{\rm n}}$: $(a=0.14,b=0)$, $(a=0,b=0.13)$,
$d=d'=2$, the shift is taken equal to $s_{1,2}=\pm0.67$ instead of $0.6$.
(c) Example of baseflow velocity $v_z$, $a_z=0.08$, $d=1.8$, this
maximum is set near $y=-0.2$, it is compared to velocity profiles
averaged conditionally for $L_x=110$,
$L_z=72$, $R=370$. The conditional average procedure is detailed in part $1$ \cite{isp1}.}
\label{profb1}
\end{figure}

We first demonstrate the effect of varying $a,b,d$ and $d'$ in $ah_1^d/\max_y(|h_1^d|)+bh_2^{d'}/\max_y(h_2^{d'})$ in figure~\ref{profb1} (a). Increasing $a$ or $b$ increases the amplitude of this linear combination. Increasing $d$ or $d'$ makes the respective function narrower around its maximum. Linear combination of functions $h_1$ and $h_2$ are then displayed in
figure~\ref{profb1} (b) with fitted $a,b,d$ and $d'$. They compare well to $v^{\rm p,n}$. If $d=d'$ they have the
$y\leftrightarrow -y, v\leftrightarrow -v$ centro-symmetry of the average profiles profiles. The two inverse of shear layer
thicknesses are \emph{a priori} not equal in the flow.  The parameters $s_1\equiv0.6$ and $s_2\equiv-0.6$
  select the position of the maximum. They are fixed at $\pm0.6$.
The shear layer at the wall is represented by the $1-\exp(-0.05(1\pm
y))$ dependence. Together with $s_{1,2}$, the value $0.05$ sets the steepness of this shear
layer: it is expected to be smooth (figure~\ref{mdl}). The thickness of this shear layer varies slowly with $R$. It is correlated to that at $y=0$, measured by $1/d$. Both are limited by viscosity.

  The baseflow, including the laminar PCF contribution is written
\begin{equation}\bar{V}_x(y)=y+\frac{ah_1^{d}(y)}{\max_{y}(|h_1^{d}(y)|)}+\frac{bh_2^{d'}(y)}{\max_{y}(h_2^{d'}(y))}\,.\label{bgf}\end{equation}
Parameters of that model will be sampled from DNS. Hereafter, we will term $ah_1$ the flow, with its amplitude $a$, and $bh_2$, the backflow, with its amplitude $b$.

Eventually, one can compute the wall normal average of $\bar{V}_x$
\begin{equation}
\frac{1}{2}\int_{-1}^1\bar{V}_x {\rm d}y\simeq -0.4(a-b)\,.
\label{inty}
\end{equation}
This will be of interest when we compare the group velocity of the instability to the wall normal average of the flow.

\subsection{Sampling the parameters of the baseflow\label{param}}

  In order to perform the stability analysis on realistic profiles, we must sample the parameters of the functional model from the DNS. For that matter we use the fact that for each $x$ position, the statistical state of the streaks is invariant along the diagonal $z'$ direction. We will therefore need only one distributions of $a,b,d,d'$ at each position $x$. These parameters come in distributions and not as single values because, however frozen or slowly evolving the turbulent streaks are, there are still fluctuations from one position $z'$ to another.

\begin{figure}
\centerline{\includegraphics[height=5cm,clip]{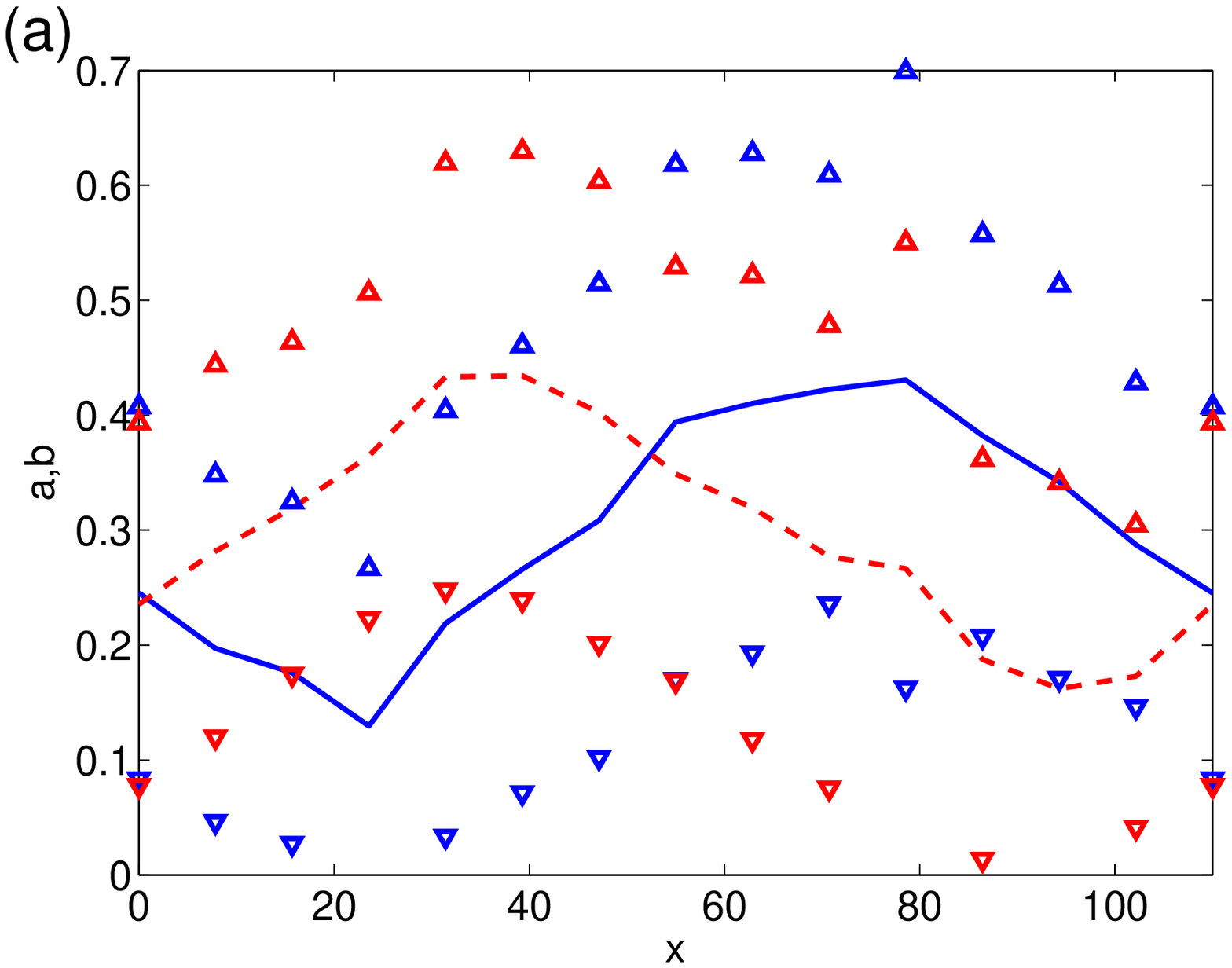}\includegraphics[height=5cm,clip]{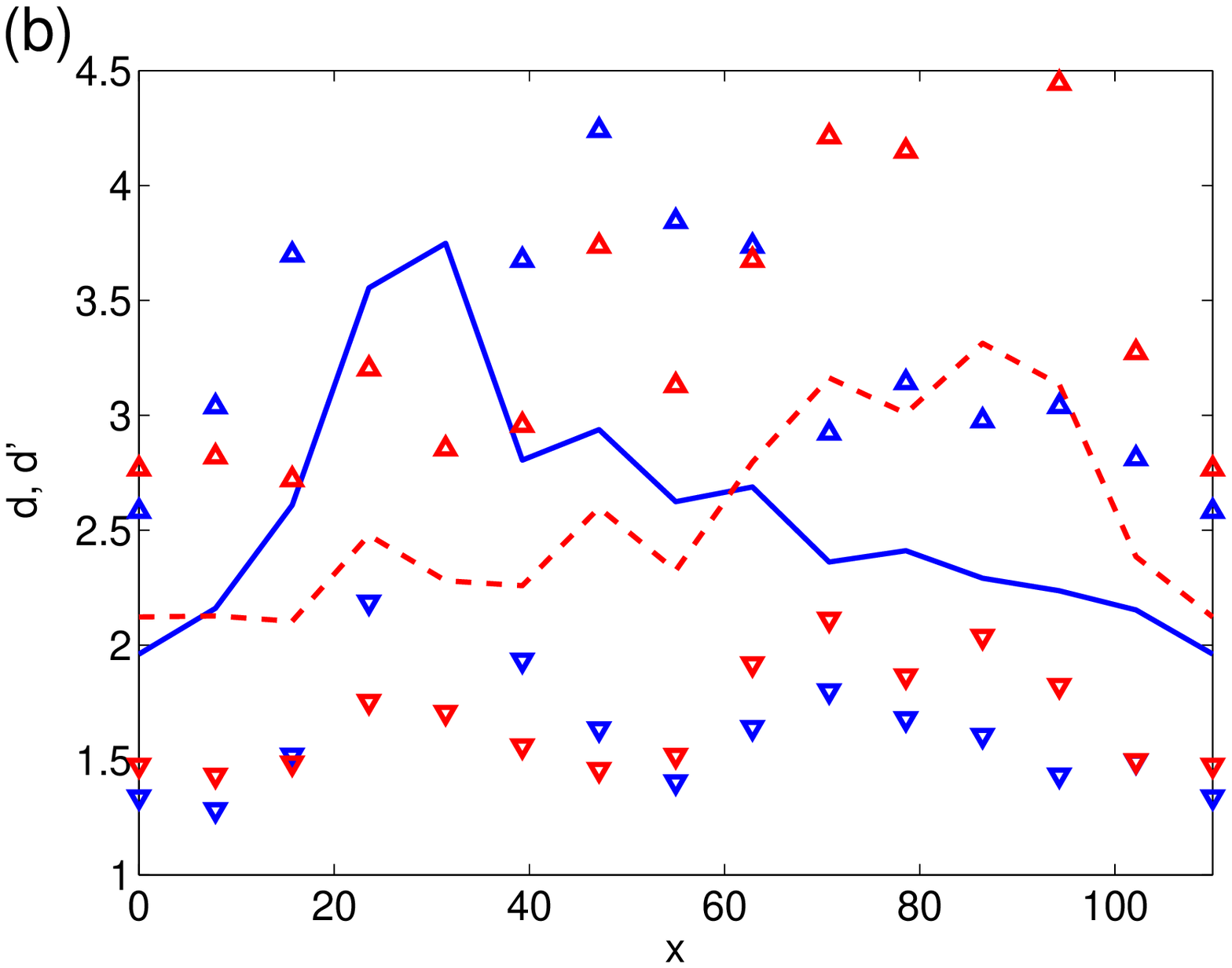}}
\caption{Fit of time averaged streamwise velocity
profiles. The continuous line represents the average, the upward (resp. downward) triangles show average plus standard deviation (resp. average minus standard deviation). (a): Amplitude $a$ (blue, flow) and $b$ (red, backflow) as a function of position
$x$. (b): Inverse of shear layer thickness $d$ (blue, flow) and $d'$ (red, backflow) as a function of
position $x$. }\label{profb2}
\end{figure}

In order to perform the sampling, we use the following method. The oblique turbulent bands are obtained in DNS by our procedure. We use  $L_x\times L_z=110\times 72$, $R=370$. The velocity field is integrated over a duration $T=25h/U$ a duration  long enough to smooth out the various irregularities, but short enough to keep the spanwise modulation of the flow. Indeed part one showed that the flow remained sinusoidal and coherent in the $z'$ direction for averaging durations of more than $100h/U$ \cite{isp1}, while Barkley \& Tuckerman showed that a duration of $2000h/U$ was necessary to completely average out the small-scale modulation \cite{BT07}. Then, at each $x$ position, the function $ah_1^d/\max_y(|h_1^d|)+bh_2^{d'}/\max_y(h_2^{d'})$ is fitted against the departure from the laminar flow over the whole range of $z'$. This produces a distribution of $a,b,d,d'$ for each position $x$. The averages $m_{a,b,d,d'}(x)$ and standard deviations $\sigma_{a,b,d,d'}(x)$ are used to characterise each distribution. We display $m_{a,b,d,d'}(x)$ (full lines) and $m_{a,b,d,d'}(x)\pm\sigma_{a,b,d,d'}(x)$ (triangles) as functions of $x$ in figure~\ref{profb2} (a,b).

  The amplitudes vary quasi-sinusoidally with a wavelength of $110$, as expected from the description of equation~\ref{eqmod2} (Fig.~\ref{f1} (b,c)). The average of the amplitude $m_{a,b}$ is proportional to the modulation of turbulence. The maxima of $m_{a,b}$ corresponds to the intermediate zone. The large crossover value (average at $\simeq 0.3$) corresponds to the middle of the turbulent zone. The small crossover value (average at $\simeq 0.2$) corresponds to the laminar zone. The standard $\sigma_{a,b}$ deviation corresponds to the short wavelength $2h$ modulation of amplitude of the turbulent streaks. It depends weakly on $x$. One typically finds an amplitude of $0.7$ in the core of the streaks, as indicated by $m_{a,b}+\sigma_{a,b}$. The turbulent fluctuations contribute weakly to the standard deviation. If one follows the approach of Shimizu \& Kida of a shear layer \cite{SK}, $m+\sigma$ is the amplitude one must choose.

The inverse shear layer thicknesses $d,d'$ depend weakly on $x$ (Fig.~\ref{profb2} (b)), it typically fluctuates between $1$ and $4$.
One can see that the amplitude of $d$ and $d'$ are not correlated. One does not expect $d=d'$ at a given $z'$ position in the baseflow: this can be understood as a consequence of the $z\leftrightarrow z+\lambda_z/2$ part of the centro-symmetry at small scale (the shift of the turbulent streaks, see Fig.~\ref{figalpha}). The equality in the average profiles $d=d'$ is reintroduced by averaging.


\subsection{A model of Reynolds stress \label{mrs}}

  In this section we propose a model of Reynolds stress which accounts for the force that maintains the slow variation of the profiles with $x$. Indeed, the slow non-parallel profiles are not maintained by a force known \emph{a priori} that is taken in the steady Navier--Stokes equations. It is the result of the turbulent activity: the self sustaining process and additional feedback mechanisms \cite{SK}. It should account for the fact that the slowly evolving baseflow sees the viscosity $(1/R)\Delta \overrightarrow{\bar{V}}_x$  balanced by advection $(\vec \bar{v} \cdot \vec \nabla)\overrightarrow{\bar{V}}_x$. the inability of eddy viscosity to balance profiles of the type of $v^{\rm p,n}$, as well as the shortcomings of $k-\omega$ type models of closure \cite{BDT} calls for an \emph{ad hoc} type of modeling. We propose here a model of Reynolds stress which is adapted to the profiles and the regime of Reynolds number. We Start from the fact that the base flow is balanced mainly by advection by the large scale flow and streamwise vortices. One can propose a stress of the type $\Sigma(y) \bar{V}_x$, with $\Sigma$ deriving from  $ \bar{V}_x(2\pi)/L_x+\omega_x $, $\omega_x$ representing the streamwise vortices. Using this multiplicative scalar assumes that the stress is ``diagonal''. Then $\Sigma(y)$ verifies at each location in the flow
  \begin{equation}
  \Sigma\bar{V_x}-\frac{1}{R}\partial_y^2\bar{V}_x\simeq 0
  \end{equation}
  The stress $\Sigma(y)$ is determined numerically: for each profile $\bar{V}_x$, $\Sigma$ is the function of $y$ which minimises the left hand side.

\subsection{Long wavelength spanwise flow\label{modz}}

We eventually move to the modeling of the spanwise flow. Since the effect of the streamwise vortices is taken into account in the Reynolds stress, we only need to model the large scale flow, noted $\bar{V}_z$. For that matter, we use the same type of profile as was used to model the flow in the velocity streaks (Eq.~(\ref{profa1})): earlier studies showed that streamwise and spanwise profiles were very similar \cite{BT07,ispspot}. One then only needs to determine the streamwise modulation of amplitude and the values of inverse shear layer thickness necessary.

The inverse shear layer thickness $d$, $d'$ and the parameters $s_{1,2}$ determining the position of the maximum are extracted from average profiles found in the region intermediate between laminar and turbulent flow (Fig.~\ref{profb1} (c)). The profiles are computed by the same method as the vorticity profiles of part $1$: we use a laminar/turbulent discrimination that allows us to determine the position $x,z$ corresponding to pseudo-laminar flow, turbulent flow and to the two intermediate regions \cite{isp1}. The spanwise flow is then conditionally averaged: this procedure averages out the small scale oscillations and therefore the contribution of the streamwise vortices and only retains the contribution of the large scale flow. The profiles conditionally averaged in the intermediate zones are well fitted by functions $h_{1,2}^{d,d'}$ (Eq.~(\ref{profa1}),~(\ref{profa2})), this leads to $s_{1,2}=\pm0.2$ and $d\simeq d'\simeq 1.8$. The amplitude $a_z$, $b_z$ of the spanwise component of the large scale flow depends on the angle \cite{BT07} and is typically contained in the range $[0,0.08]$.

Our analysis can be applied to several configurations of laminar-turbulent coexistence of plane Couette flow which display different configurations of the spanwise large scale flow. The simplest case is that of the oblique bands. The spanwise large scale flow is sinusoidal, and we can perform the same type of rewriting of the result of Barkley \& Tuckerman \cite{BT07}, putting the stress on
the flow in the intermediate area (like in equation~\ref{eqmod2})
\begin{equation}  \bar{V}_z=v_z^{\rm p}\left(1+\cos\left(2\pi \left(\frac{x}{L_x} + \frac{z}{L_z}\right) \right)\right)+v_z^{\rm n}\left(1+\cos\left(2\pi \left(\frac{x}{L_x} + \frac{z}{L_z}\right)+\psi \right)\right)\,.\label{bgfz}\end{equation}
This leads to a spanwise large scale flow well described by $h_1$ or $h_2$ alone in the intermediate regions.

  The case of the expanding spots can be considered \cite{ispspot}. Below $R_{\rm t}$, the spots have a steady band configuration at long time, and they are described by the former model. Meanwhile, above $R_{\rm t}$ (and below $R_{\rm t}$ at short time), there is a quadrupolar flow around the spot. The velocity profiles of that quadrupolar flow are well described by the functions $h_{1,2}$, and our analysis can be applied to the study of the advection of small scale perturbations around the spot \cite{PRL}. The study of a linear perturbation to slowly expanding structures implies some subtleties \cite{ACC}, however, the linear study of a frozen baseflow can give a reliable first step to understand the problem.

  Eventually, one can consider the case of plane Couette flow disturbed by a purely spanwise wire \cite{DD,BT99}. In that case the laminar-turbulent coexistence displays a purely spanwise front with no spanwise large-scale flow. Linear study of the laminar baseflow around the wire showed that it became unstable after $R$ was increased above a subcritical bifurcation threshold. However, like the case of the bands, there is no information on the sustainment mechanism of the laminar-turbulent structure. The study of that case can bring insight. It is modeled by our streamwise baseflow with $\bar{V}_z=0$.

\section{Linear stability problem \label{lsp}}

\subsection{Framework\label{fw}}

In this section, we detail the linear stability problems that we solve in order to study the formation of spanwise vorticity.
We start with the case of the local (or quasi-parallel) study of a baseflow representing a location of the bands or the spots, with two components $(\bar{V}_x,\bar{V}_z)$. The three
component perturbation is denominated $\mathbf{u}$. We use the Orr--Sommerfeld--Squire
system in which the perturbation to the frozen velocity streaks is
described by the wall normal component of the vorticity
$\eta=\partial_z u_x-\partial_x u_z$ and the wall normal component
of the velocity $u_y$
\begin{equation}\partial_t \Delta u_y+\left(\bar{V}_x\partial_x+\bar{V}_z\partial_z \right)\Delta u_y -\left(\partial_y^2( \bar{V}_x)\partial_x+\partial_y^2(\bar{V}_z)\partial_z \right)u_y-\frac{1}{R}\Delta^2 u_y=0\label{oss}\,,\end{equation}
\begin{equation}\partial_t \eta+\bar{V}_x\partial_x \eta+\bar{V}_z\partial_z \eta+\left(\partial_y(\bar{V}_x)\partial_z-\partial_y(\bar{V}_z)\partial_x\right) u_y-\frac{1}{R}\left(\partial_x^2+\partial_y^2+\partial_z^2 \right) \eta=0\,,\label{sq}\end{equation}
with boundary conditions
\begin{equation}u_y(y=\pm 1)=0,\quad\partial_y u_y(y=\pm 1)=0,\quad \eta(\pm1)=0\,.\label{bc1} \end{equation}
See Schmid \& Henningson  for a derivation and a general
discussion \cite{DR}.

Using a Fourier decomposition in streamwise and spanwise decomposition, both perturbations are of the type
\begin{equation}\eta(y,x,t)=\hat{B}(y)\exp\left(\imath k_xx+\imath k_zz +\sigma t-\imath \omega t\right)+c.c.\end{equation}
and
\begin{equation}u_y(y,x,t)=\hat{C}(y)\exp\left(\imath k_xx+\imath k_zz +\sigma t-\imath\omega t\right)+c.c.\,,\end{equation}
$c.c.$ denoting the complex conjugate.
They are the eigenmodes of our eigenvalue problem. We perform a temporal
stability analysis: the wavenumbers $k_x$ and
$k_z$ are taken as parameters and the growth rate $\sigma$ and
frequency $\omega$ are the eigenvalues
\begin{align}
\partial_t (\Delta u_y,\eta)=L_{k_x,k_z}(u_y,\eta)\\ \Rightarrow (\sigma +\imath \omega)(\Delta u_y,\eta)=L_{k_x,k_z}(u_y,\eta)\,,
\end{align}
with $L$ the linear operator of equations~(\ref{oss},\ref{sq}). The group velocity $(c_x,c_z)=(d\omega/dk_x,d\omega/dk_z)$ will be computed numerically, by finite differences, once the dispersion relation $\omega(k_x,k_z)$ is known. This calculation is a fundamental step in local stability analyses.

Due to our quasi-parallel
approximation, small values of $k_x$ and $k_z$ have no physical
relevance. The
wavevector of the band is approximately $2\pi/L_x\simeq 0.05$, a lower
boundary for our analysis can be taken at $k=0.1$.

The case of a one-component baseflow is the main interest of
this article. It allows one to discuss the effect of the Reynolds stress, and the absolute or convective nature of the instability very simply. Besides, it models the case of plane Couette flow disturbed by a wire. It is described by the Orr--Sommerfeld equation~(\ref{oss}) in which $\bar{V_z}=0$
\begin{equation} \partial_t \Delta u_y-\frac{1}{R}\Delta^2 u_y +\bar{V}_x \partial_x \Delta u_y-\partial_x u_y \partial_y^2 \bar{V}_x=0\label{OSu}\,,\end{equation}
with the same boundary conditions 
\begin{equation}u_y(y=\pm 1)=0,\quad\partial_y u_y(y=\pm 1)=0\label{bc2} \end{equation}.
We use a perturbation of the form
\begin{equation}u_y(y,x,t)=\hat{A}(y)\exp\left(\imath kx +\sigma t-\imath\omega t\right)+c.c.\,,\end{equation}
$A$ is the eigenmode and $\sigma$ and $\omega$ are the eigenvalue.
Again, the lower values of $k$ have no physical relevance.
Taking into account our model of Reynolds stress, one can arrive at an equivalent to the Orr--Sommerfeld for $u_y$ equation. One replaces $\bar{V}_x\partial_x$ by $\bar{V}_x\partial_x+\Sigma$ in Navier--Stokes equations and performs the operations leading to the Orr--Sommerfeld equation. This leads to
\begin{equation}\partial_t \Delta u_y+\left((y+\bar{V}_x)\partial_x+\underbrace{\Sigma}_{\rightarrow c_g} \right)\Delta u_y-\partial_xu_y \partial_y^2(y+\bar{V}_x)+\underbrace {\partial_y\Sigma\partial_yu_y}_{\rightarrow \sigma}=\frac{1}{R}\Delta^2u_y\,.\label{eqforce}
\end{equation}
The stress modeled by $\Sigma$ appears in two terms: one impacting mainly the group velocity and one impacting mainly the growth rate. A computation of both terms show that the first one is small relative to its classical Orr--Sommerfeld counterparts, while the other one is not negligible. One can therefore expect that only the growth rate will be notably modified.

  The case of global stability analysis with a two component baseflow $\bar{V}_x$, $\bar{V}_y$ is eventually considered. The corresponding Orr--Sommerfeld equation can be derived and yields for the streamfunction $\psi$
\begin{align}
\partial_t \Delta \psi +\bar{V}_x\partial_x \Delta \psi-\partial_x \psi \partial_y^2\bar{V}_x+\bar{V}_y\partial_y \Delta \psi\left(\Delta \psi \partial_x \bar{V}_x+\partial_y \psi \partial_x\partial_y \bar{V}_x \right)+\left(\Delta \psi \partial_y \bar{V}_y-\partial_y \psi \partial_x^2\bar{V}_y+\partial_x \psi \partial_x\partial_y \bar{V}_y \right)=\frac{1}{R}\Delta^2\psi\,.
\label{eqglob}
\end{align}
Note that $\bar{V}_x$ depends on $x$, and that a second component $\bar{V}_y$ is necessary to maintain incompressibility.
The eigenmode is a function of $x$ and $y$. It is decomposed on
Fourier modes in the streamwise direction.

\subsection{Numerical procedure \label{num}}

  The numerical procedure uses two bases of orthonormal polynomial fitting
the boundary conditions~\ref{bc1} and~\ref{bc2}. It has been used for low-order modeling of the transition, or a simple description of the rolls in part $1$ \cite{isp1}. It has also been used in an attempt to obtain a numerical model of wall normal turbulence with lowered numerical cost \cite{lm}. This description originates from the study of thermal convection \cite{conv}. This procedure allows a
simple implementation of boundary conditions and quick convergence. Further details on this
type of numerical approach can be found in the
literature \cite{chqz,phd}.

\begin{figure}\centerline{\includegraphics[width=7cm]{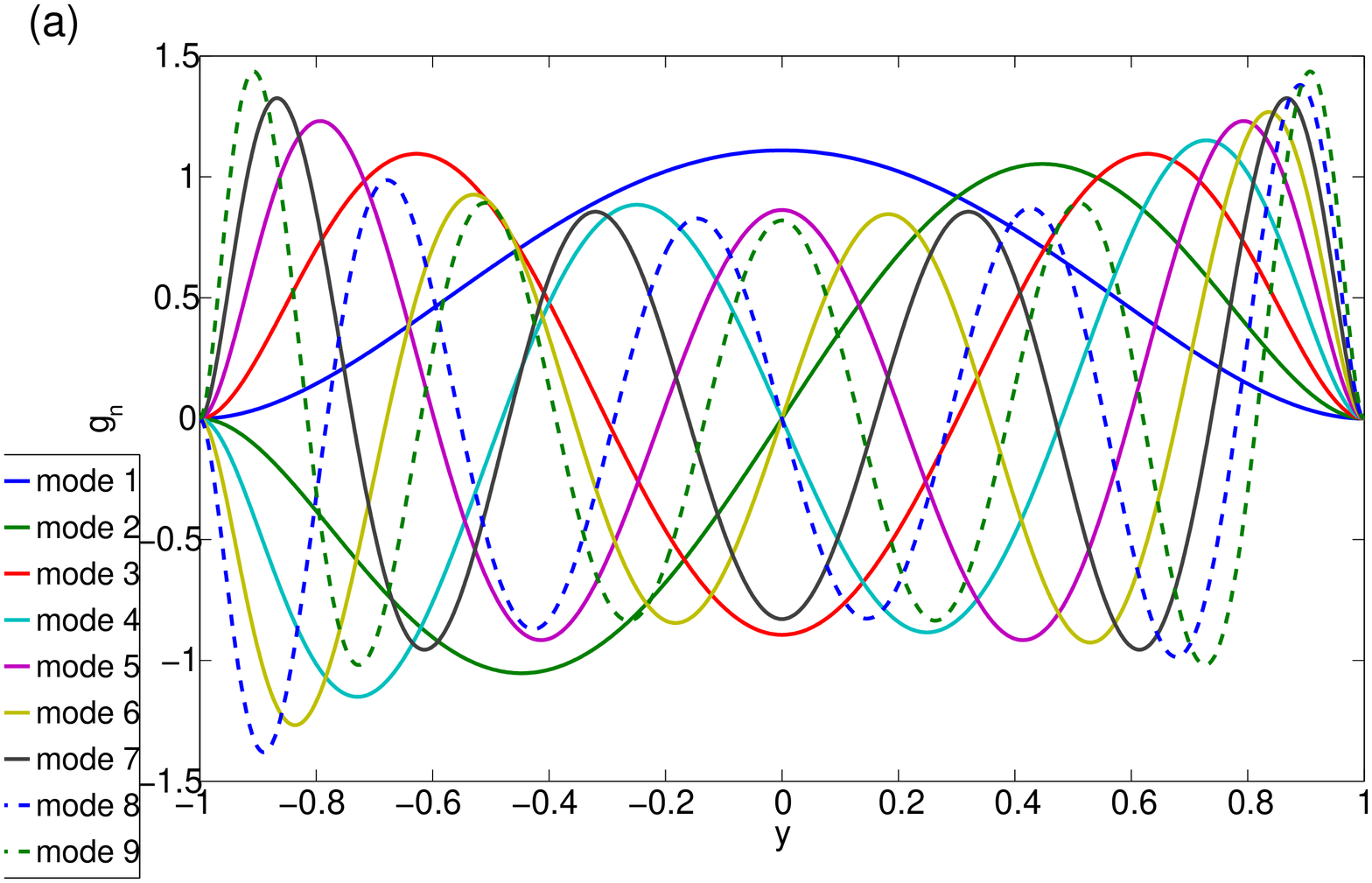}\includegraphics[width=7cm]{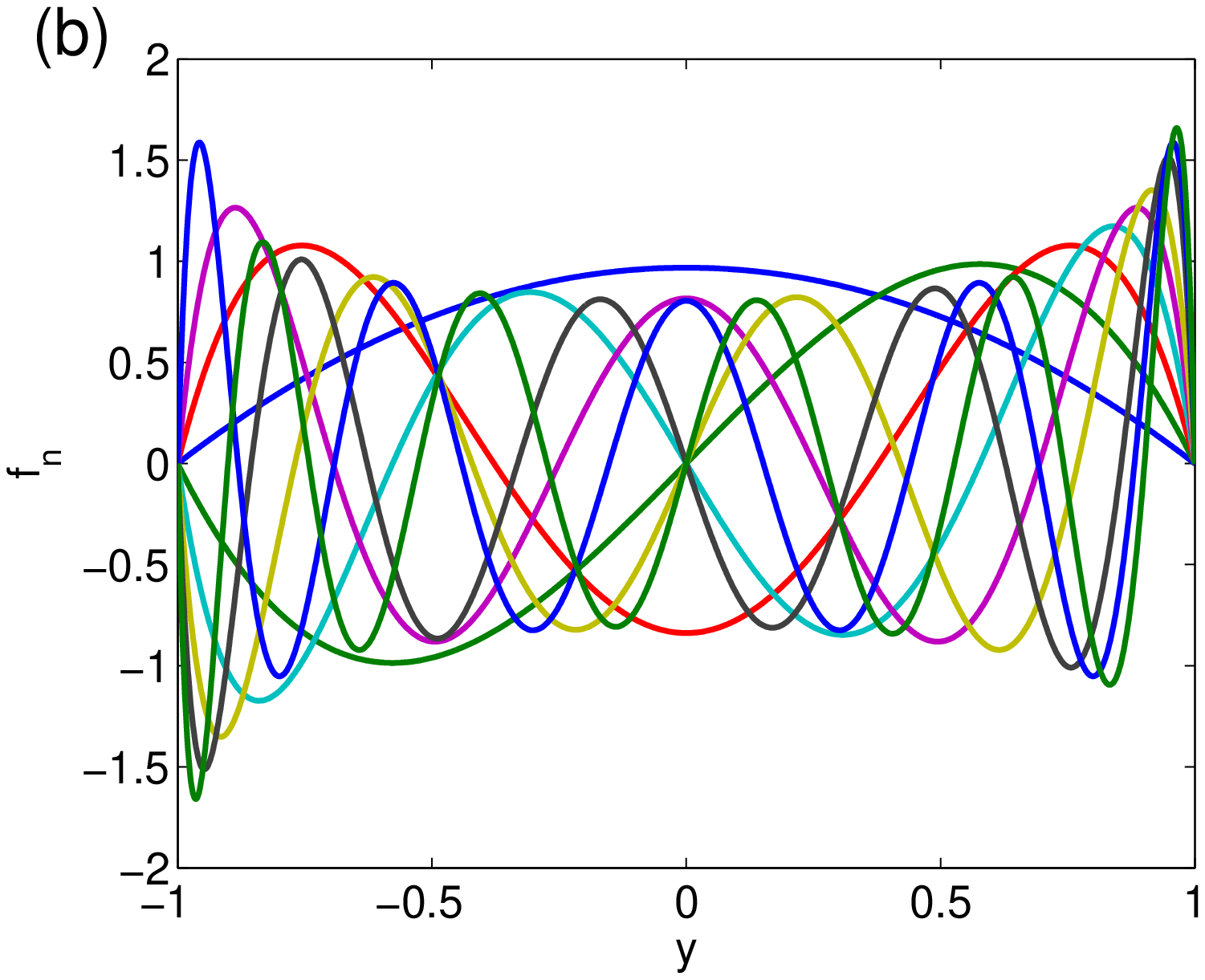}}
\caption{(a) example of $g_n$ polynomials. (b) example of $f_n$
Polynomials}\label{fpol}
\end{figure}

The basis $\{g_k \}$ is used for $u_y$ and $\phi$: both the functions and their first derivative go to zero at $y=\pm1$. The basis $\{ f_l\}$ is used for $\eta$: this function goes to zero at $y=\pm1$. The first two normalised polynomials of each basis are
\begin{equation}g_0=\frac{\sqrt{315}}{16}(1-y^2)^2\,,\, f_0=\frac{\sqrt{15}}{4}(1-y^2)\,.\end{equation}
The next functions can be written as
\begin{equation}g_k=P_k(y)(1-y^2)^2\,,\, f_l=Q_l(1-y^2)\,. \end{equation}
The polynomials $P_k$ and $Q_l$ are determined \emph{via} a
Gram-Schmidt orthonormalisation process (see \cite{linalg} for instance). These polynomials
correspond to $2,2$ (for $Q_l$) and $4,4$ (for $P_k$) Jacobi
polynomials \cite{chqz} multiplied respectively by $(1-y^2)$ and $(1-y^2)^2$. However, the formulation obtained from the Gram--Schmid
process is more tractable. The first polynomials of both bases can
be seen in figure~\ref{fpol}.

A Galerkin truncation at order $n$ for $u_y$ and $n'$ for $\eta$ is used to obtained
the algebraic eigenvalue problem. The function $\hat{A}(y)$ and $\hat{C}(y)$ are then developed and truncated
\begin{equation}\hat{A}(y)= \sum_{k=0}^n a_kg_k,\quad \hat{C}(y)= \sum_{k=0}^n c_kg_k \,,\end{equation}
while $\hat{B}(y)$ is developed and truncated 
\begin{equation} \sum_{l=0}^{n'} b_lf_l\,.\end{equation}
Following the Galerkin Procedure, the
equation~\ref{oss} and~\ref{OSu} are integrated against $\{g_k\}$
and the equation~\ref{sq} is integrated against $\{g_k\}$, ($k\le
n$) and $\{f_k\}$ ($k\le n'$). The integrals up to order $n$ are
computed using collocation points and the corresponding
weights \cite{chqz}. The roots of the polynomial $g_{2n+2}$ are the collocation points when integrating $g$ polynomials. They are denoted by $\{y_k^g\}$.
In the same manner, the roots of the polynomial $f_{2n'+2}$ are the collocation points when integrating $f$ polynomials. They are denoted by $\{y_l^f\}$. The corresponding weights $\{\rho_k^g \}$ and
$\{\rho_l^f\}$ are determined for exact polynomial
integration \cite{chqz}. They are the solution of the systems
\begin{equation}\int_{y=-1}^{1}{\rm d}y\,y^{2m}\left(1-y^2\right)^4 =2\sum_{k=1}^{n+1}\rho_k^g
\left(y^g_k\right)^{2m}\left(1-\left(y_k^g\right)^2\right)^4 \end{equation} and
\begin{equation}\notag\int_{y=-1}^{1}{\rm d}y\,y^{2m}\left(1-y^2\right)^2=2\sum_{l=1}^{n'+1}\rho_l^g
\left(y_l^f\right)^{2m}\left(1-\left(y_l^f\right)^2\right)^4\,.
\end{equation}

  Convergence of the method was checked on a hyperbolic tangent profile with
walls, for various shear layer thicknesses \cite{phd}. The growth rate is the most sensitive on resolution. In our range of Reynolds numbers, precise results are obtained for $n\gtrsim 15$. The number of modes used for precise results in the study is $n=n'=24$.

\section{Linear stability analysis\label{res}}

  This section explores the characteristics of the instability. The one-component local case is considered first, to present the main mechanisms of the instability, and a comparison with the phenomenological study. The convective/absolute character of the instability is considered on that case. We then move to the local two components case, to explore the effect of the spanwise sweep. Eventually a global case is considered, for qualitative insight and comparison with the local study.

\subsection{One-component base flow, effect of shape\label{res1}}

  This case is described by Orr--Sommerfeld equation (\ref{OSu}). Temporal linear stability is performed. Two typical cases are considered in the first subsection, to illustrate the features of the instability. A parametric study, including the effect of the Reynolds number, is performed in the second subsection.

\subsubsection{General properties}

  The examples of profiles studied here correspond to
the typical intermediate intermediate zone with $\bar{V}_x<0$ in the top part ($b=0$, $a= 0.7$, $d=3.9$, $b=0.7$, $a= 0$, $d'=3.9$) and to the typical turbulent zone ( $a= 0.7$, $b =0.3 $, $d'=3$) at $R=350$, in agreement with DNS data (figure~\ref{profb2},
(a,b)). They illustrates the main
characteristics of the two dimensional instability. In both cases, the analysis is performed with and without our model of Reynolds stress to explore its effects.

\begin{figure}
\centerline{\includegraphics[height=5cm,clip]{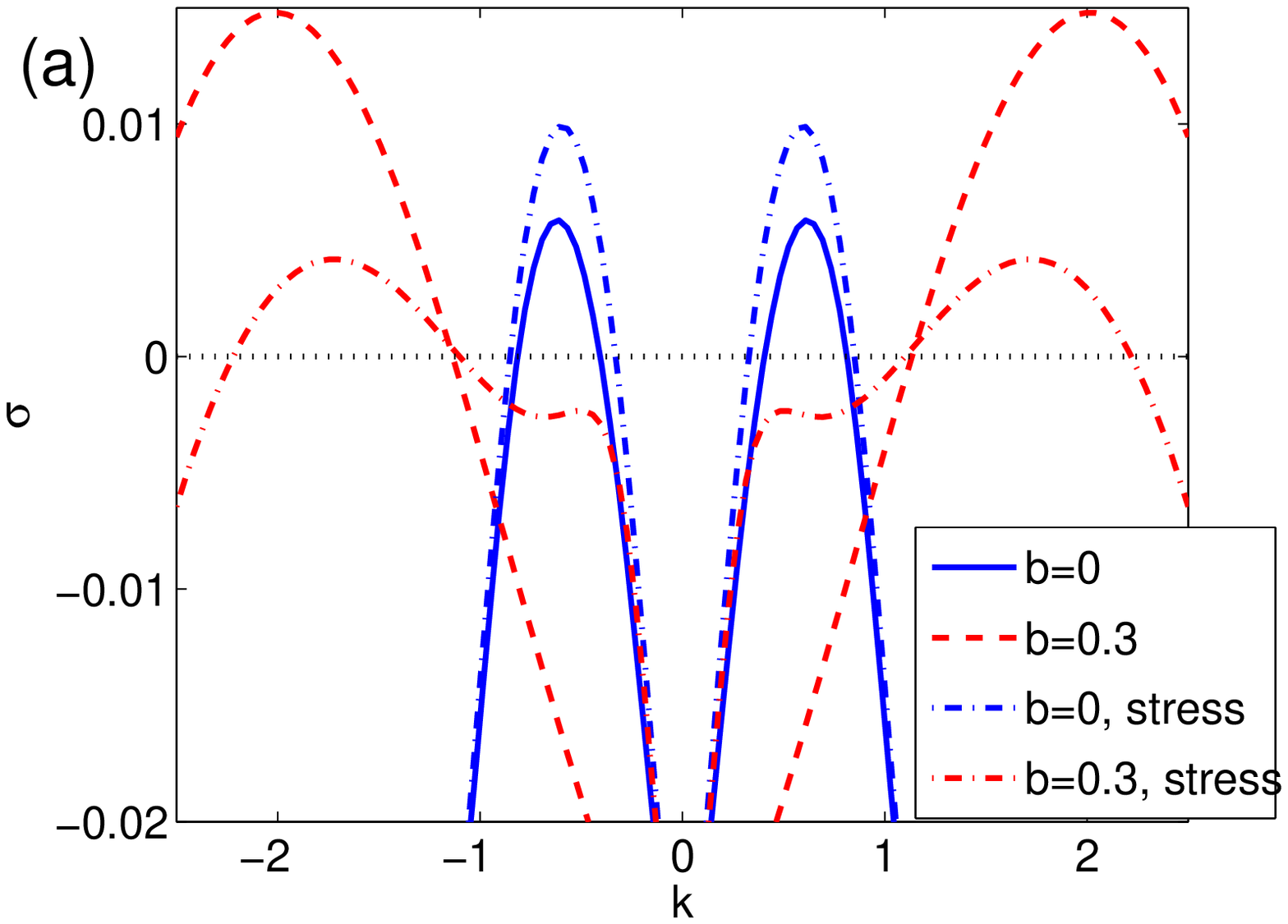}\includegraphics[height=5cm]{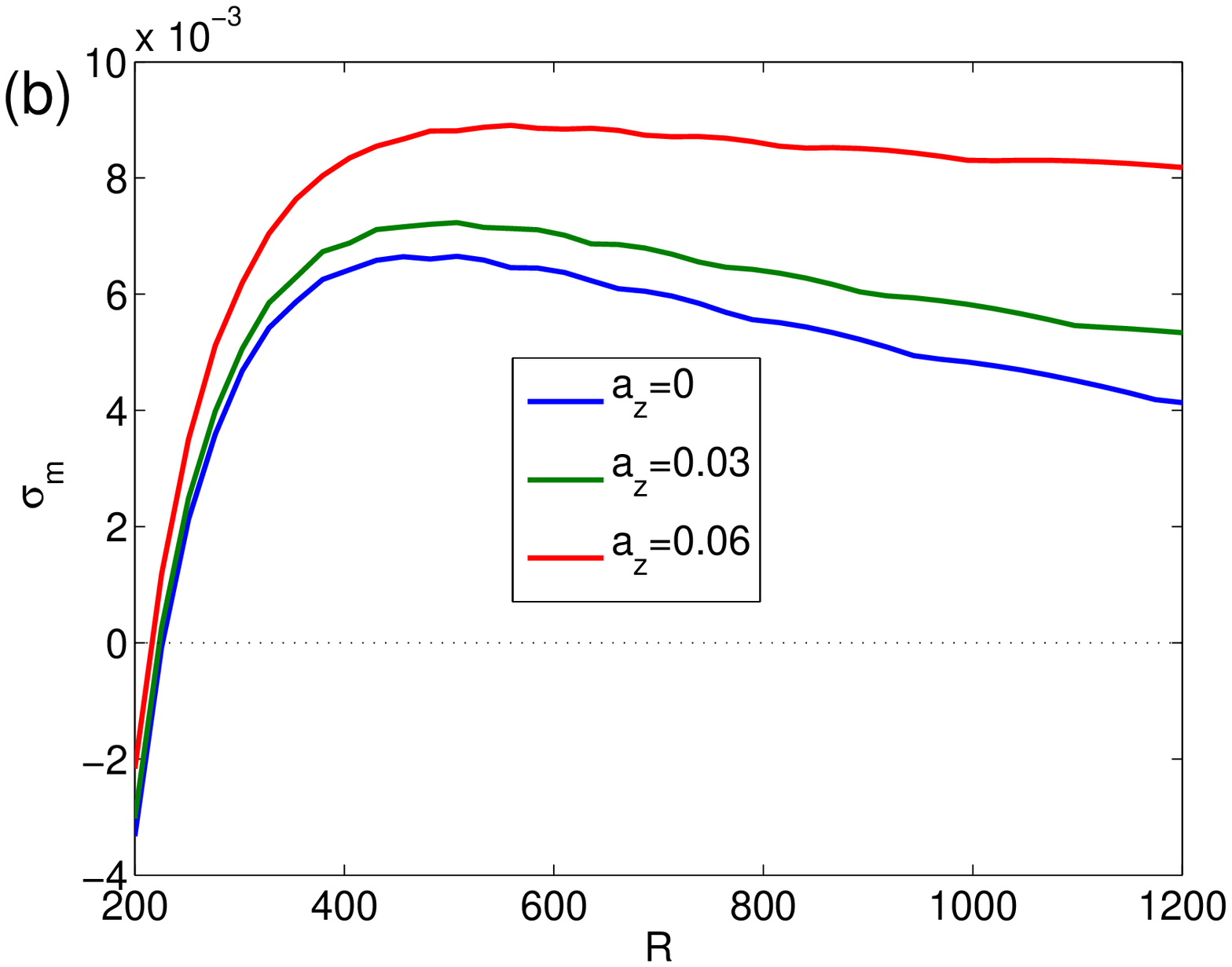}\includegraphics[height=5cm,clip]{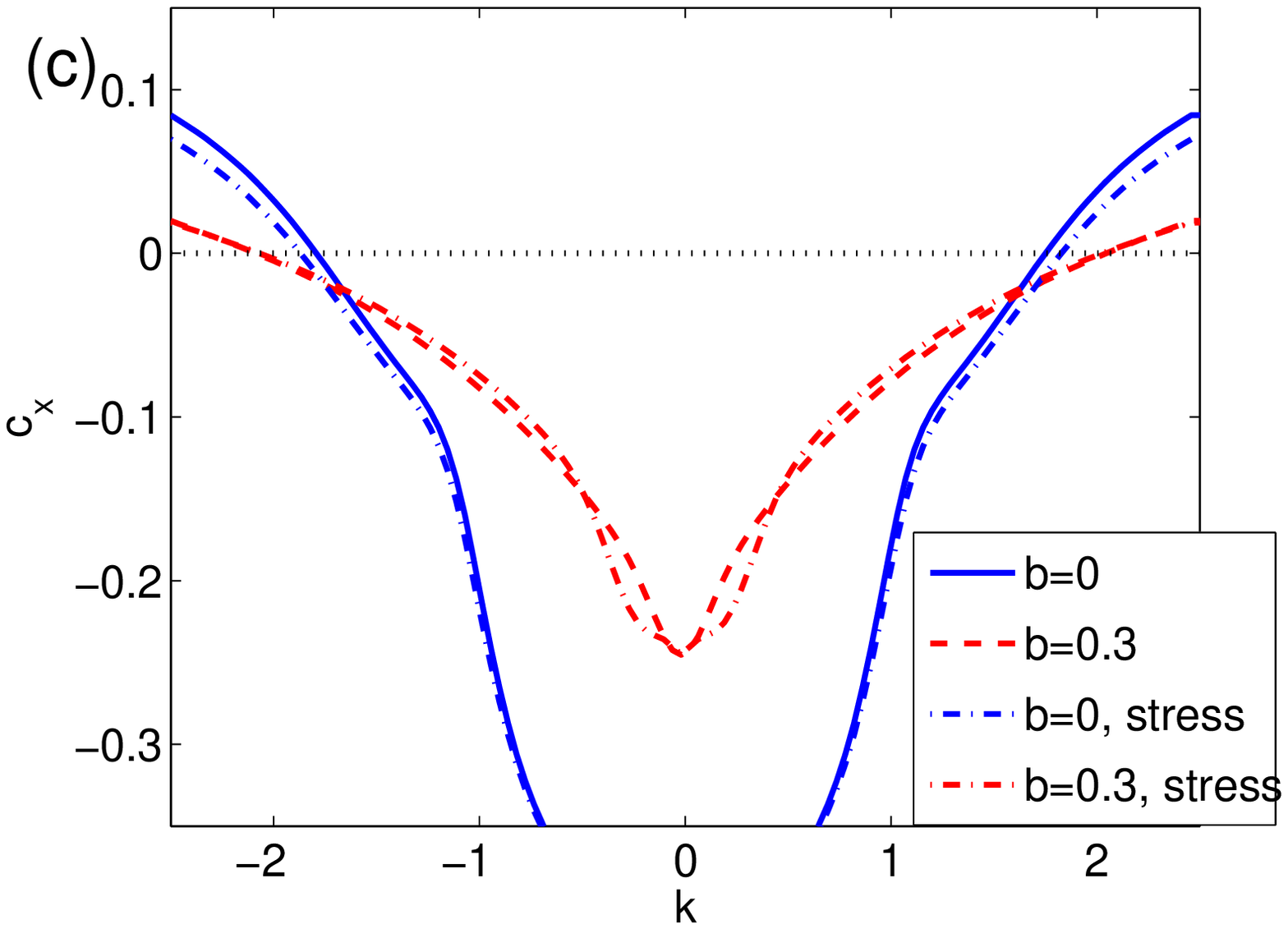}}
\caption{(a) growth rate vs $k$ for $a=0.7$, $d=3.9$, $b=0$ and
$b=0.3$ $d'=3$, $R=350$, plus points for $a=0$, $b=0.7$, $d'=3.9$,
$R=350$. (b) maximum of $\sigma(k)$ vs $R$, (same parameters) for three values of $a_z$, $d_z=1.8$.(c)
group velocity vs $k$ centered around the $\sigma>0$ wavenumbers
(same parameters $R=500$)}\label{artres}
\end{figure}

The growth rate for the two profiles is displayed as a function of
the wavenumber, in figure~\ref{artres} (a). One finds the typical
behaviour of $\sigma$ for a viscous Kelvin--Helmholtz instability: $\sigma$ depends only on $|k|$. There is a positive
maximum for $\sigma$ at $|k|=O(1)$ and purely viscous decrease of
$\sigma(|k|)$ at large wavenumbers. Application of the
$y\leftrightarrow -y$, $v_x\leftrightarrow -v_x$ symmetry does not
change $\sigma$. The parameter $b$ (and
therefore the amplitude of the backflow) increases the most unstable wavenumber from $k_c(b=0)\simeq0.5$ and $k_c(b=0.3)\simeq2$ (Fig.~\ref{artres} (a)). Both are of order $1$ and increase as one enters the turbulent zone, in agreement with the DNS. The quasi-parallel
approximation is self-consistent, since both  wavenumbers are much larger than that of the band. Eventually, we compare the results of the analysis with and without stress. One can see that the Reynolds stress can increase or decrease the growth rate, depending on the situation, however, neither the stability, nor the position of the maximum is changed

The maximum of $\sigma(k)$ is denoted $\sigma_m$. It is found at $\pm k_c$. It is displayed vs the Reynolds
number in figure~\ref{artres} (b). Only the $a_z=0$ plot is of
interest at this point: the three curves are discussed together in
the next section. Viscosity stabilises the base flow for the
smaller values of $R$. The
growth rate $\sigma$ then crosses $0$ at $R_s\sim200$, the threshold Reynolds number of the instability, reaches a maximum at
$R_m\sim400$, then decreases toward a plateau value at larger $R$. It reaches the inviscid regime, which is outside of the range of Reynolds number of interest. The Reynolds stress has a weak effect on $\sigma_m(R)$, it is not displayed here.

In order to discuss the advection of the perturbations, we turn to the group velocity. In the framework of absolute and convective instabilities \cite{HM}, one can ask two questions. Firstly: in the longtime limit, is the linear response going to grow where it appeared (absolutely unstable) or be advected away (convectively unstable) ? Secondly: if it is advected, at what speed does the corresponding wavepacket move ?

\begin{figure}
\centerline{{\Large \textbf{(a)}\includegraphics[width=7cm,clip]{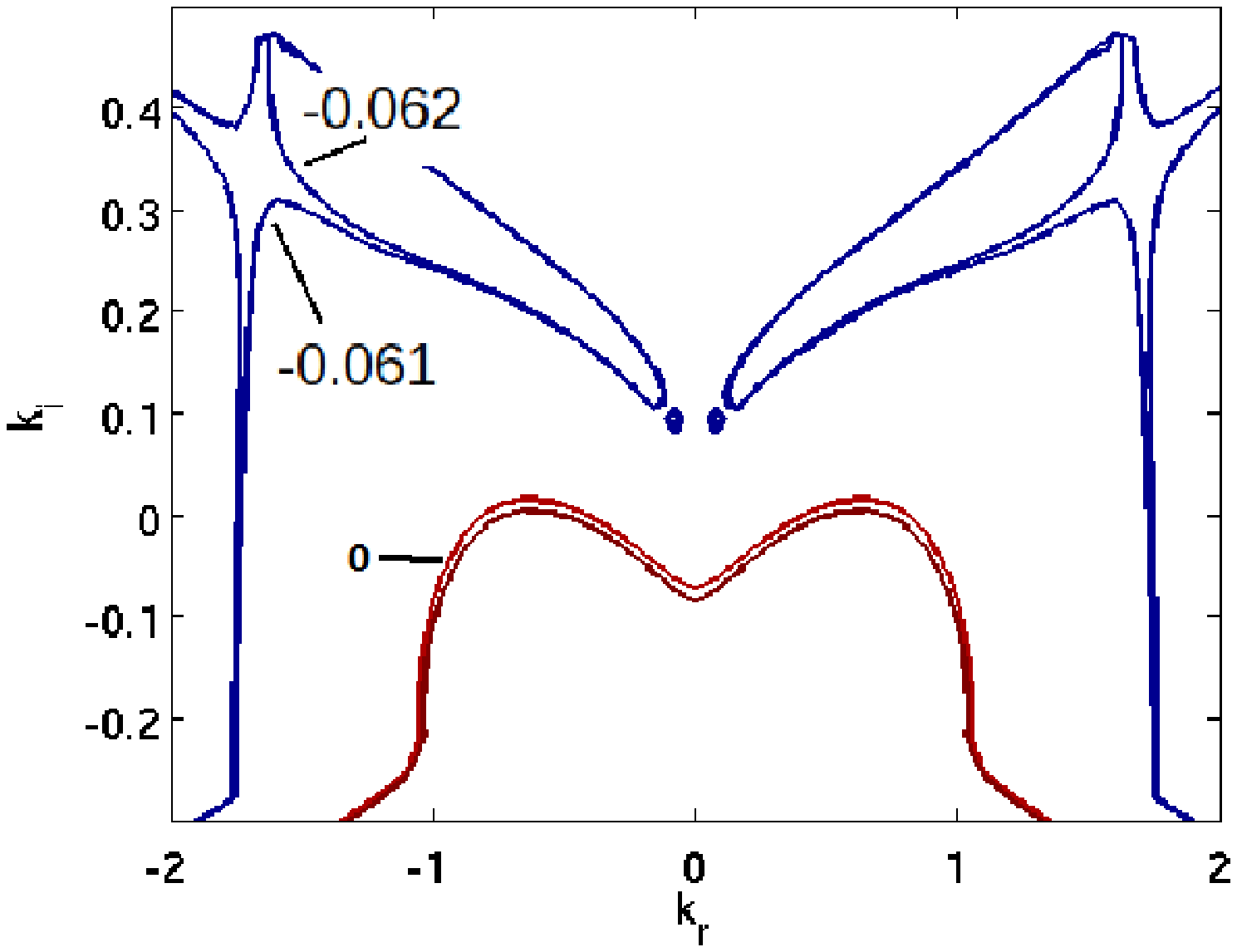}}{\Large \textbf{(b)}\includegraphics[width=7cm,clip]{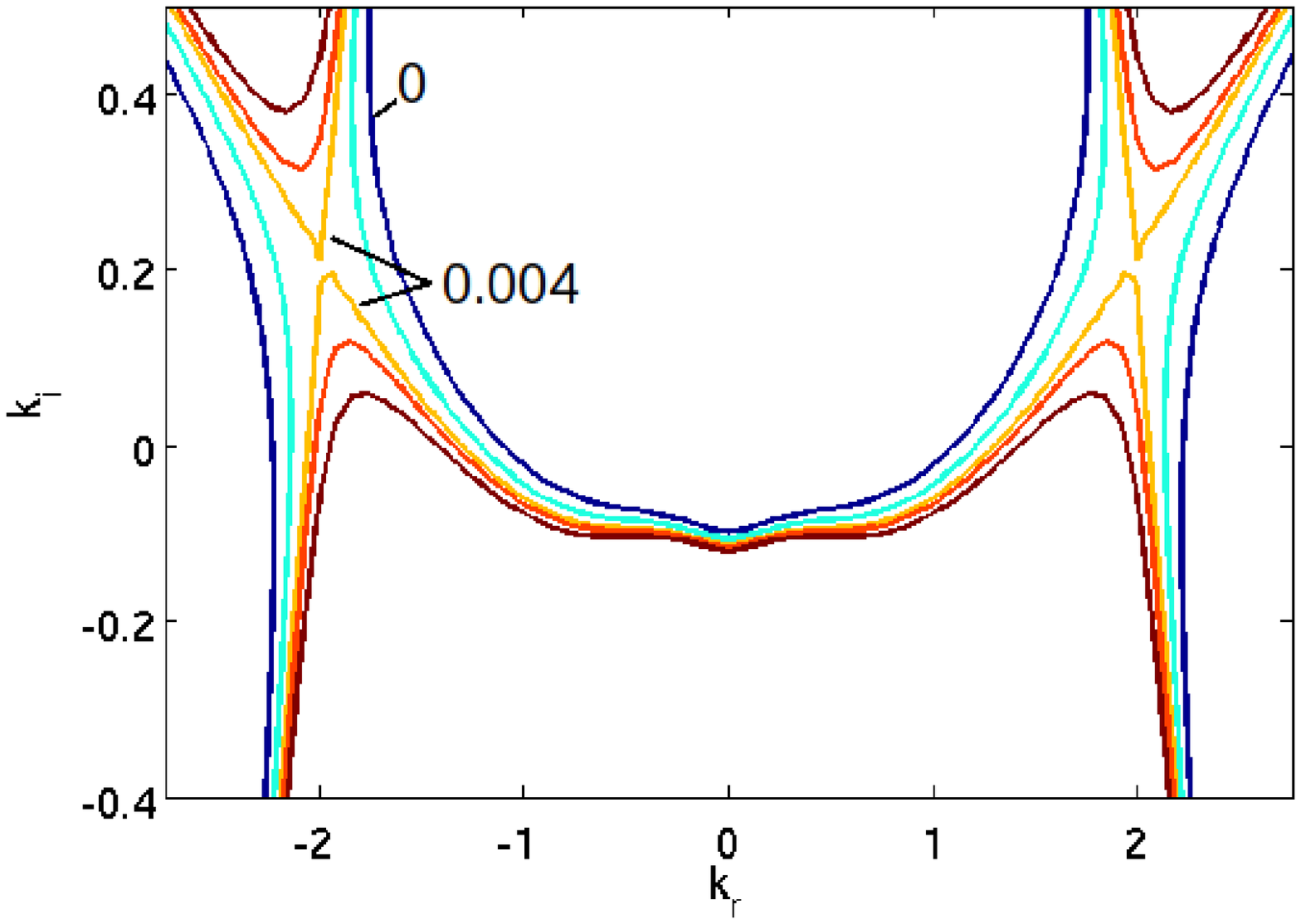}}}
\caption{Isovalues of the growth rate $\sigma$ in the complex plane $k_i,k_r$, resulting from the stability analysis with $a=0.7$ and $d=3.9$. (a): Result without any backflow $b=0$, typical of the intermediate zone. (b): result with backflow, $b=0.2$, $d'=3$, typical of the turbulent zone.}
\label{plcp}
\end{figure}

In order to answer the first question, one needs to compute the growth rate at the zero of the group velocity. This particular growth rate is denoted $\sigma_0$. This choice follows the study of the linear impulse response problem, which yields the Green function of the problem $G(x,t)$ \cite{HM}. All linear initial value problems can indeed be written as a convolution of this Green function. In the long time limit, $G$ can be calculated using a saddle point approximation. This calculation depends on the ray $x/t=\text{cte}$ we are considering, and it yields that $G$ grows or decays like $\exp(\sigma_{x/t}t)$. Then, in order to determine whether the instability is absolute, one need to consider the saddle point calculation in which we assume that the perturbation remains at $x=0\Leftrightarrow x/t=0$ (where the impulse originally was), which gives the zero group velocity condition. In order to consider the growth rate alone, one can then use the Briggs--Bers criterion, which uses the properties of the complex dispersion relation $\omega(k)$. The locus of the zero group velocity corresponds to pinch point of the $k^+$ and $k^-$ branches of $\sigma$ to determine the position of that point in the complex plane. This is applied to our two examples without (Fig.~\ref{plcp} (a)) and with (Fig.~\ref{plcp} (b)) backflow. The growth rate $\sigma_0$ is negative in the case without backflow, that case is convectively unstable, whereas $\sigma_0$ is positive in the case with backflow, that case is absolutely unstable. A computation at $a=0$ and $b=0.7$ shows that the convective Vs. absolute nature of the instability is unchanged.

In order to answer the second question, one need to compute the group velocity $c_x$ at $\sigma_m$. It is found on the real axis (Fig.~\ref{artres} (c)). In the convectively unstable case of $b=0$, one finds $c_x\simeq 0.3$, which is slightly larger than the wall normal average of $\bar{V}_x$ of order $0.28$ (Eq.~\ref{inty}). A computation at $a=0$ and $b=0.7$ shows that the sign of $c_x$ is opposite, while the absolute value is the same. Eventually, we note that, as expected from the qualitative analysis of equation~\ref{eqforce}, the use of a Reynolds stress impacts the group velocity very weakly.

\begin{figure}
\centerline{\includegraphics[height=5cm,clip]{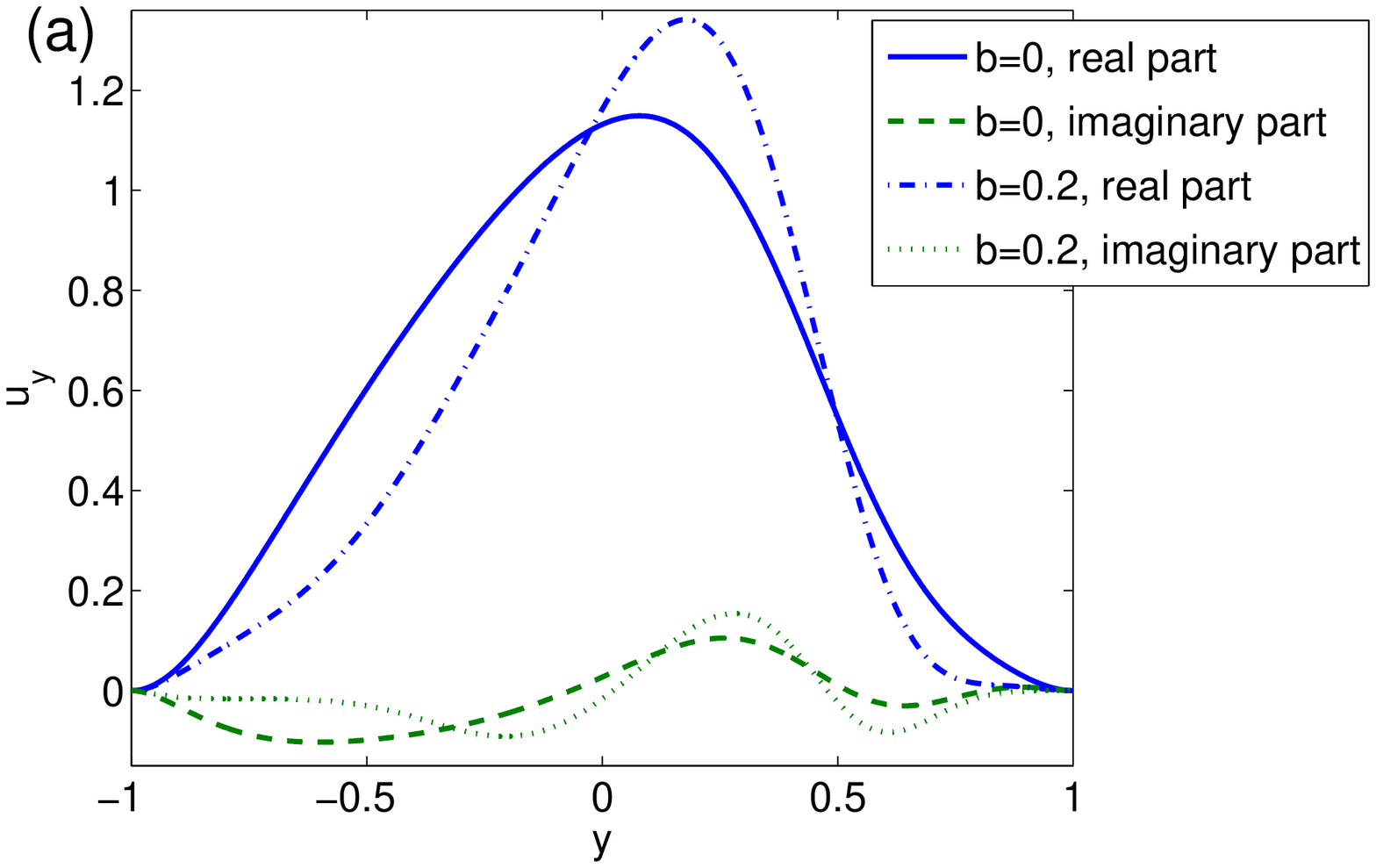}\includegraphics[height=5cm,clip]{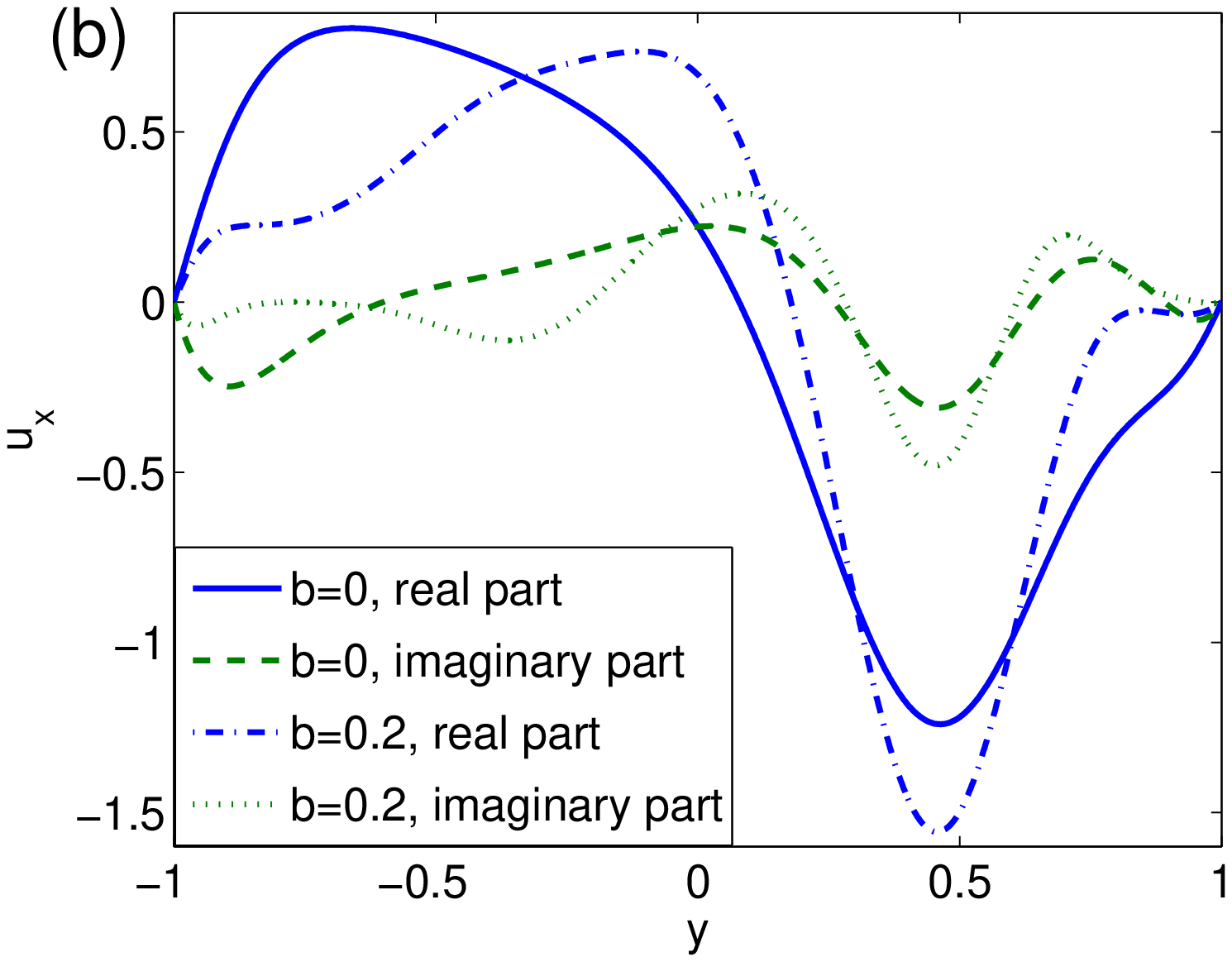}}
\caption{(a): Unstable eigenmode (real and imaginary parts)
for $R=300$, $d=3.9$, $a=0.7$, $b=0$. (b): $y$ derivative of the unstable eigenmode.}
\label{mdp}
\end{figure}

A normalised eigenmode (real and imaginary parts, $u_y$ and $u_x$)
is plotted vs $y$ in figure~\ref{mdp} (a,b). The $y$ derivative
of this eigenmode, corresponding to the unstable streamwise
component $u_x\propto i\partial_y u_y/k$ is displayed in figure~
\ref{mdp} (b). There is a dominant part and smaller sinusoidal part
dephased by $\pi/2$. The dominant part is of most interest here. In the first part of this article, it has been shown that
the departure to the velocity streaks which lead to spanwise vorticity was well describe by $f_1$ (Fig.~\ref{fpol} (b)), in the case of $u_x$, and $g_0$ (Fig.~\ref{fpol} (a)), in the case of $u_y$.
This is perfectly consistent with the results of the linear analysis: the real part of the unstable eigenmode $u_y$ (which is the largest) is well described by $g_0$ (Fig.~\ref{mdp} (a)), while the dominant part of the corresponding $u_x$ is well described by $f_1$ (Fig.~\ref{mdp} (b)).
 The presence of a
backflow $b=0.2$ does not change qualitatively the shape of the
eigenmodes. Quantitatively, their spatial scale of variation are smaller. There maxima are shifted toward the $y>0$
part of the flow.

\subsubsection{Parametric dependence}

\begin{figure}
\centerline{\includegraphics[width=6cm,clip]{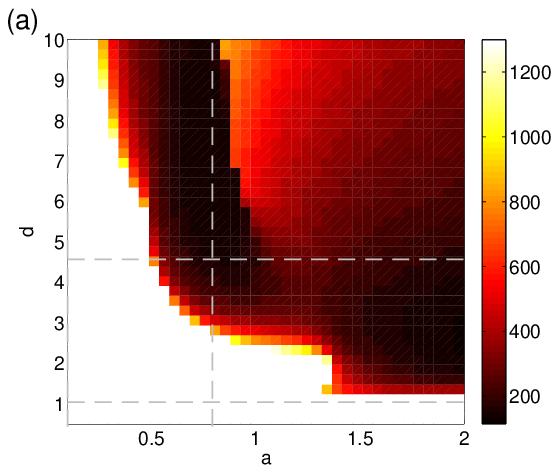}\includegraphics[width=6cm,clip]{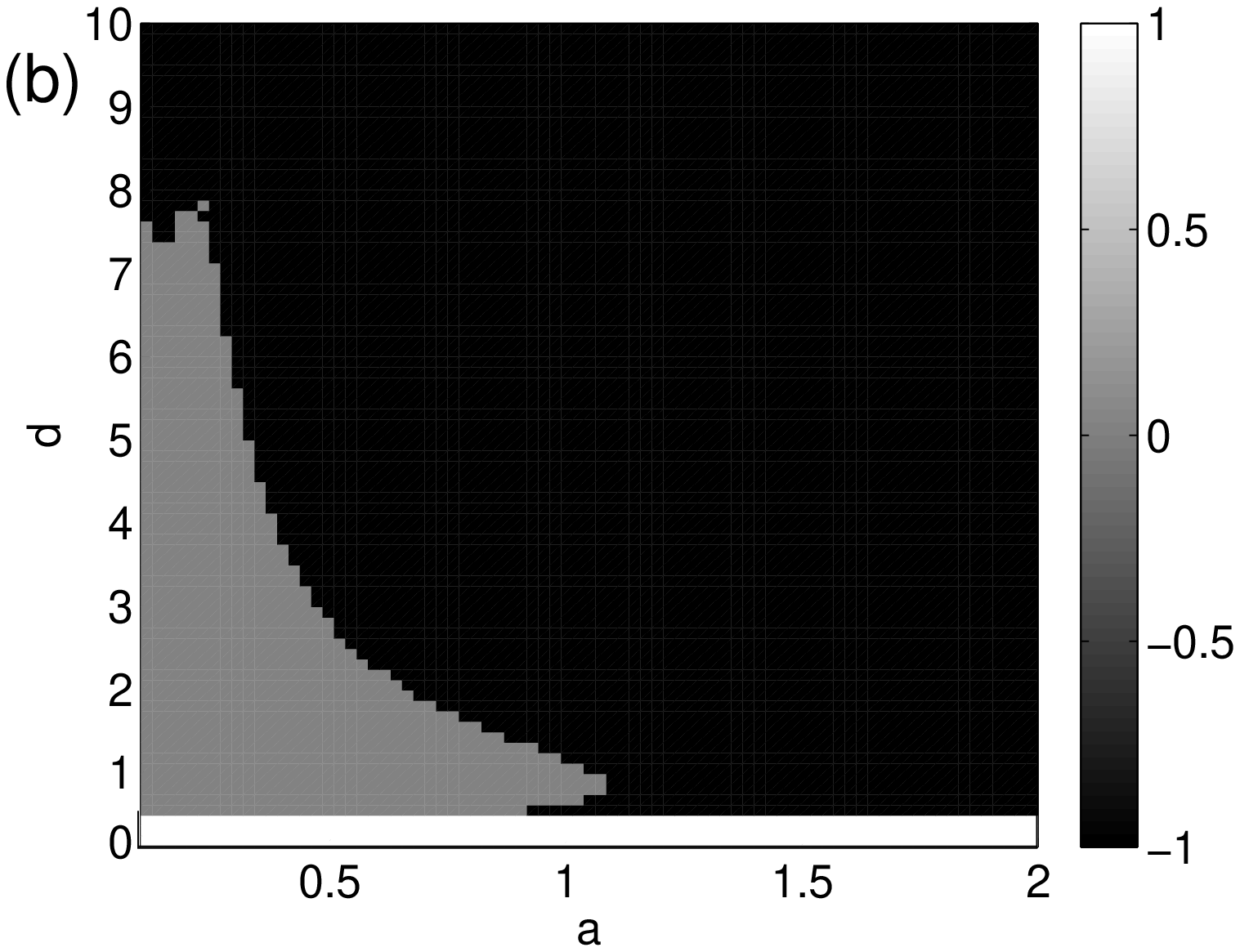}}
\caption{(a): colour levels of $R_s$ vs $a$ and $d$. The dashed gray lines indicate the range of $a$ and $d$ admissible, according to the DNS sampling (Fig.~\ref{profb2}). (b): colour plot of Rayleigh and Fj\o rtoft
criteria for this type of modelled profiles (white : no inflection
point, gray: Fj\o rtoft criterion not verified, black: Fj\o rtoft
criterion verified}\label{sm}
\end{figure}

 The threshold Reynolds number $R_s$, defined in the former section
(Fig.~\ref{artres} (b)), is displayed in colour levels figure~\ref{sm} (a) as a function of $a$ and $d$ in the case $b=0$ (intermediate zone). Values of threshold Reynolds over $800$ are only reach if $R_s$ displays a very fast increase which is possibly a divergence. No calculations have been performed above $R_s=1000$. The region of the parameter space $R_s\ge 1000$ is stable for any Reynolds number relevant for the band regime, so that the value of $R$ is not pushed any further. The smaller the amplitude $a$
or inverse shear layer thickness $d$, the larger the instability
threshold $R_s$. The same profiles can be tested against inviscid necessary criteria \cite{DR} for instability,
namely Rayleigh criterion, which requires an inflection point of the profile (at $y_s$), and Fj\o rtoft
criterion, which requires $(\bar{V}_x(y)-\bar{V}_x(y_s))\partial_y^2\bar{V}_x$ (or $\partial_y\bar{V}_x\partial_y^3\bar{V}_x$) to be negative somewhere in the flow (Fig.~\ref{profb2} (a)).
Note that both formulations of the Fj\o rtoft criterion include either $V$ or $V'$, and therefore the laminar baseflow $y$ or its derivative $1$ (respectively). The inviscid necessary criterion for stability
therefore imposes conditions on the relative amplitude $a$ and $b$ beside of the shape of the profile.
Unless viscous instabilities of Tollmien--Schlichting (TS) type occur,
the instability is expected to occur inside the area where the criteria are valid (black area). The TS instability is certainly not expected in Couette flow. Moreover, the presence of velocity streaks strongly reduces the TS instability \cite{cb}.
The area indicated by the Fj\o rtoft inviscid necessary criterion as stable is strictly included in the
stable area. The instability only develops for values of the
parameters for which an inviscid instability is allowed. This
confirms that it is of the Kelvin-Helmholtz type. The laminar flow causes the baseflow to be stable in an area much larger than that predicted by the Rayleigh criterion. Without the
laminar baseflow, the profiles introduced in equation~\ref{profa1}
or extracted from the DNS (Fig.~\ref{mdl}) would be unstable for all
$a$. It would only necessitate an inflection point to exist, \emph{i.e.} an inverse of shear layer thickness $d\gtrsim0.5$.

  The averaged profiles, found in equation~\ref{eqmod2} correspond to
small values of $a$ and $d$. One finds approximately for this type
of profiles $a\lesssim 0.3$ and $d\lesssim 2$. This range of
parameters $a,d$ is in the middle of the stable area
(Fig.~\ref{profb2} (a), Fig.~\ref{sm} (a)). This confirms that the
phenomenon observed in DNS is not an instability of an average
profile. Values of $d$ found in the flow (Fig.~\ref{profb2} (b)) are
approximately bounded by $d\lesssim 4.5$. Given this bound the
amplitude $a$ has to be large enough for the profile to be unstable
(Fig.~\ref{sm} (a)). One finds $a\gtrsim 0.5$. This range of value
of $a$ corresponds to the maximal value found in the flow
(figure~\ref{profb2} (b)). These values are only found in the core of
the velocity streaks (Fig.~\ref{figalpha} (a), Fig.~\ref{profb2} (b)), as discussed in section~\ref{param}.

\begin{figure}
\centerline{\includegraphics[width=6.5cm,clip]{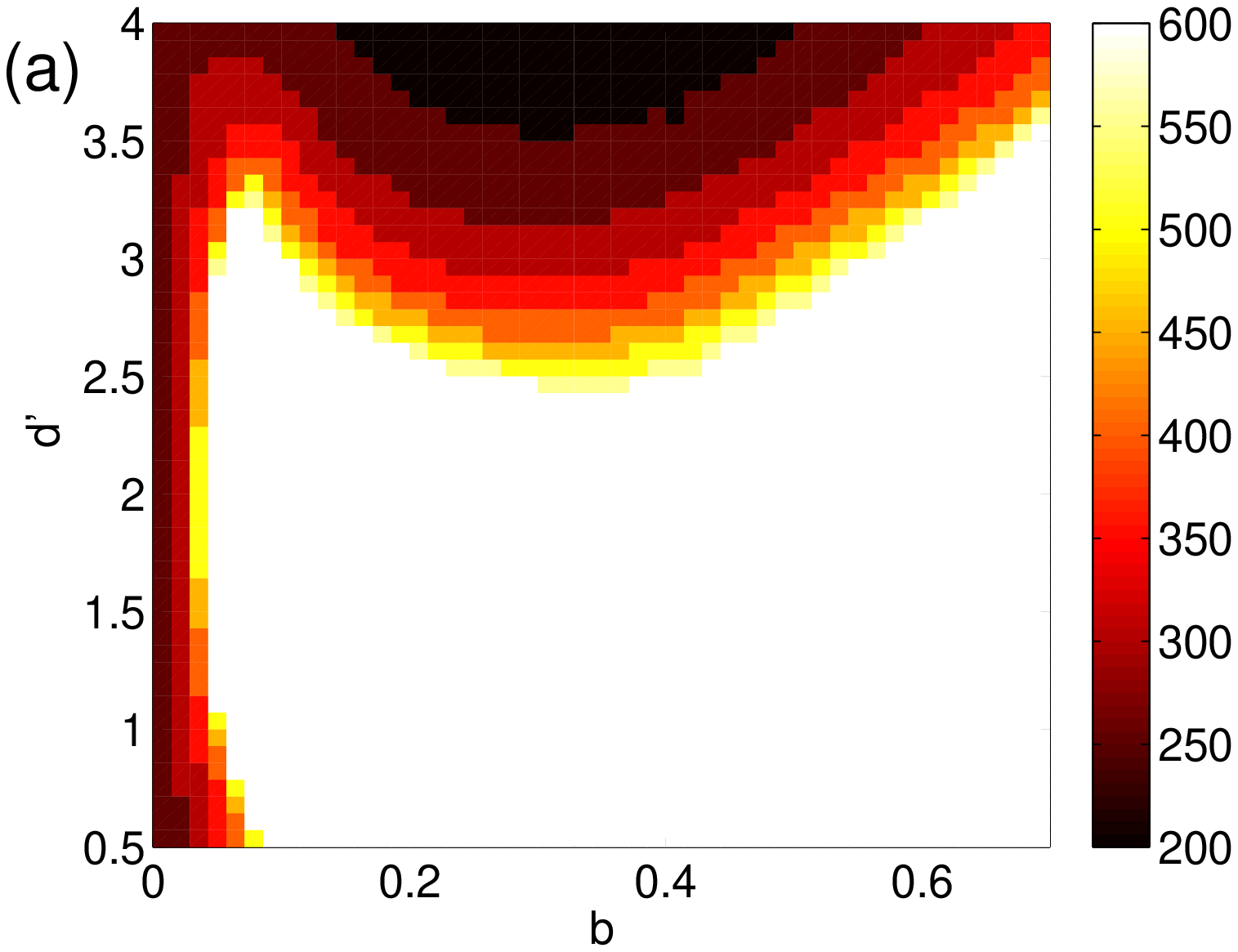}\includegraphics[width=6.5cm,clip]{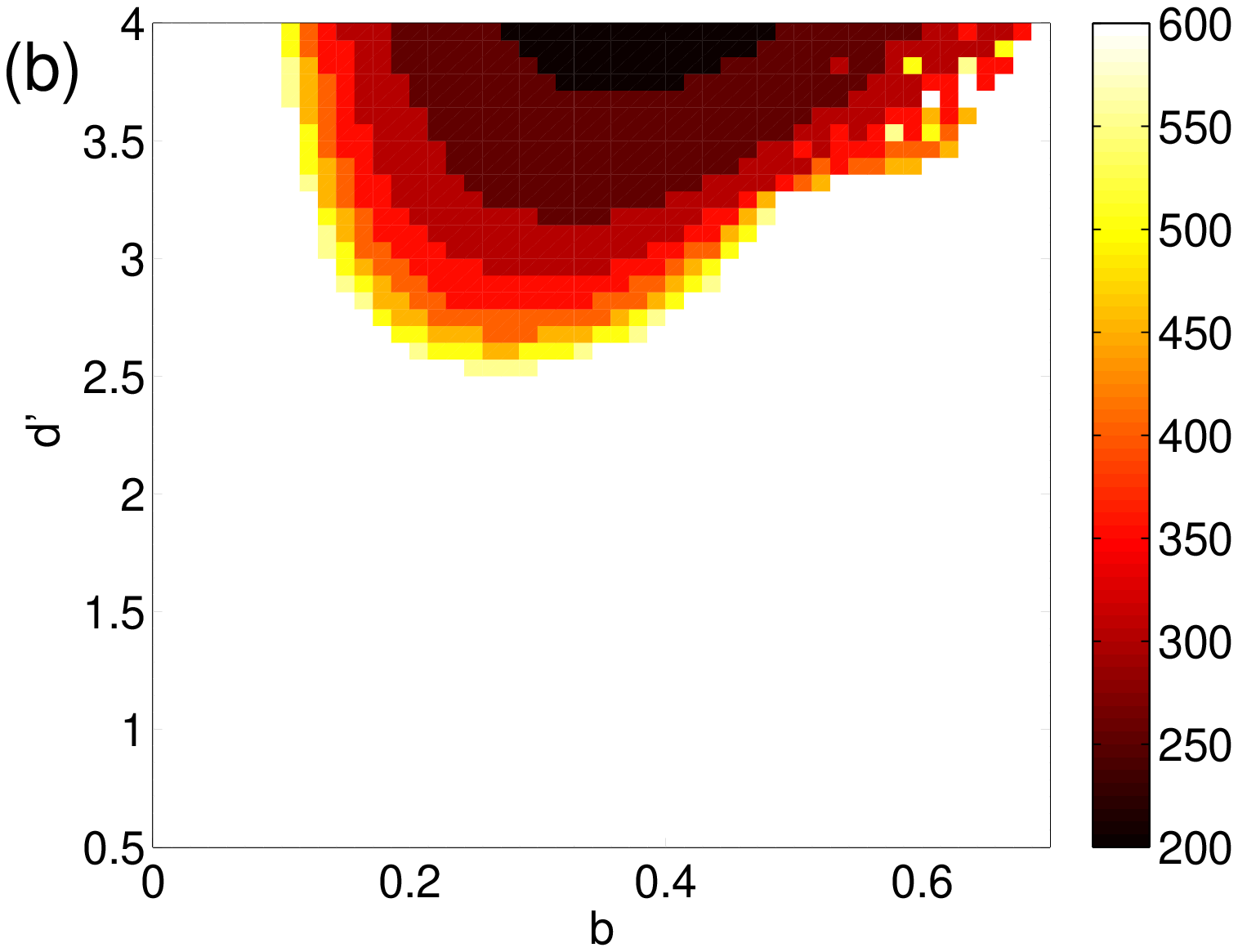}\includegraphics[width=6.5cm,clip]{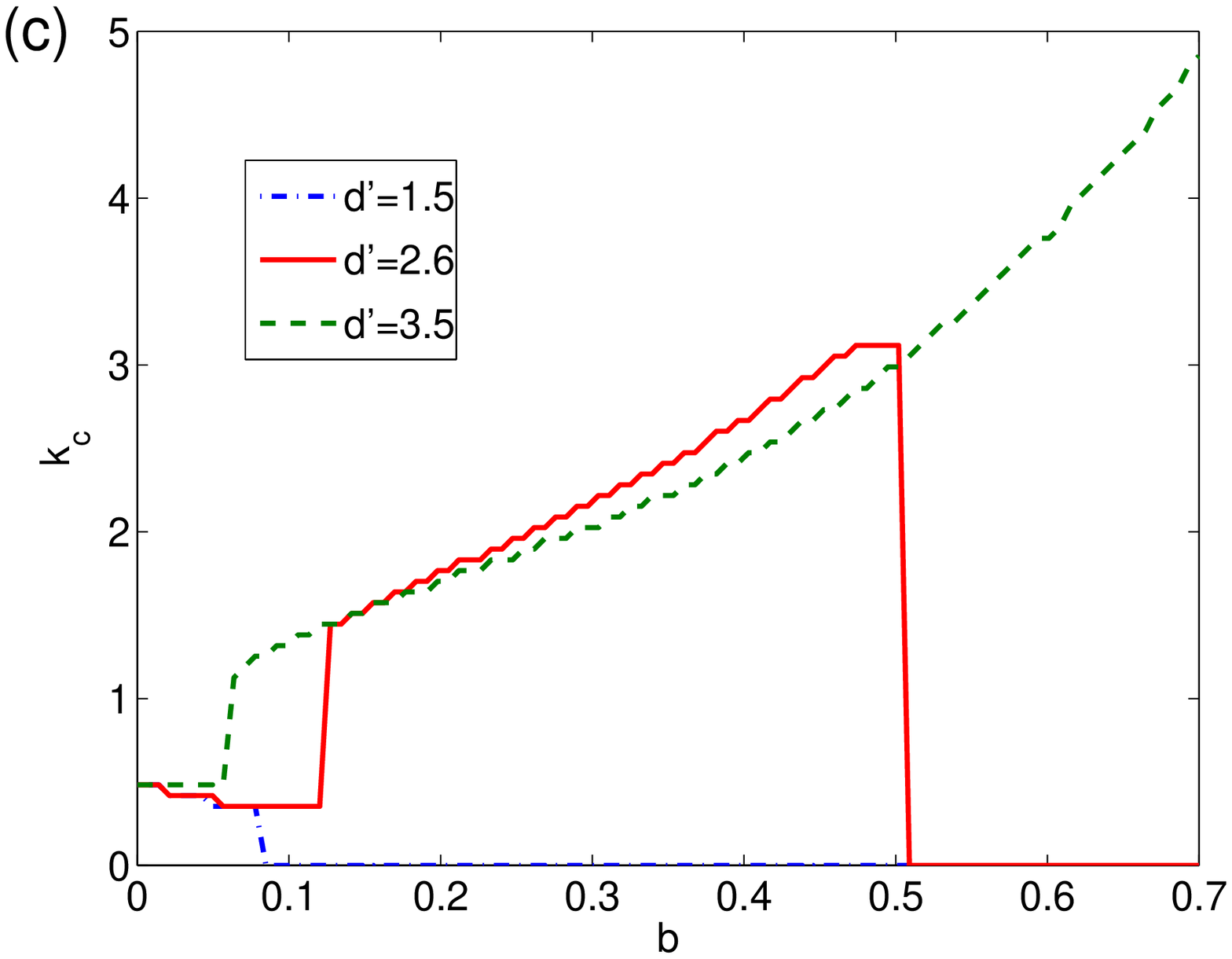}}
\caption{Stability analysis for various $b$ and $d'$, $a=0.7$, $d=3.9$.
(a) Region of the parameter space for which the flow is unstable,
(b) Region of the parameter space for which the flow is absolutely unstable for
various Reynolds numbers in the range of interest $R\in [200:500]$. (c): Maximum wavenumber as a function of $b$ for
several $d'$, $R=350$, $a=0.7$, $d=3.9$.
If $\sigma(k)$ has no maximum, $k_c$ is set to zero.}
\label{sm_}
\end{figure}

 A systematic investigation of the effect of the backflow on $\sigma_m$ and $\sigma_0$ is
performed by exploring the $b,d'$ parameter plane. This study is performed for the Reynolds number $R$
ranging from $R=200$ to $R=550$. This range contains the bands \cite{RM,isp1}, the low Reynolds number spots \cite{ispspot} and excited turbulence \cite{DD}. The amplitude and inverse shear layer thickness have the
same value as before, $a=0.7$, $d=3.9$. The sign of $\sigma_m$ and
$\sigma_0$ is computed as a function of $b$ and $d'$ for each Reynolds number. The range of
parameters for which the flow is unstable at a given Reynolds number
is displayed in the colour levels of figure~\ref{sm_}. Each colour level delimits the zone of the space of parameters inside which the flow is unstable, at the corresponding Reynolds number.  White corresponds to no instability. Figure~\ref{sm_} (a) indicates whether the baseflow is unstable at a given Reynolds number. Figure~\ref{sm_} (b) indicates whether the baseflow is absolutely unstable at a given Reynolds number. For smaller values of $d'\lesssim
2.5$, the backflow stabilises the baseflow. One finds that
$\sigma_m$ goes to zero as the line $b\simeq 0.05$ is crossed. We then consider the range $0.1\le b\le 0.3$, $2.5\le d'\le 4$: in parameter space the boundary $d'(b)$ between stable and unstable flow goes toward smaller and smaller $d'$. As a consequence, as
$b$ is increased, the flow is destabilised again. The flow can be absolutely unstable
provided $b\gtrsim0.15$ and $d'\gtrsim 2.5$ (Fig.~\ref{sm_} (b)). For the largest values of backflow amplitude, the flow is stabilised again. However, situations
where $a\simeq b$ are less likely to occur, even in the turbulent
zone. Decreasing the Reynolds number
reduces the range of parameter for which the flow is unstable. The
convective instability disappears below $R\simeq 200$. The flow is
no more unstable without backflow. The flow can be absolutely
unstable down to an even smaller Reynolds numbers $R\gtrsim 150$.
This requires a backflow and a small enough shear layer thickness.

  For a given value of $a$, there is a minimum value of $b$, depending weakly
on $R$, for which absolute instability can occur. Since the baseflow is the sum of the laminar baseflow, the flow and the backflow,
one cannot factor out the amplitude $a$ of the flow and study a normalise case. As a consequence, the minimal value $b(a)$ for which an absolute instability can occur cannot be expressed as a ratio. Note however that this is the
equivalent of the minimum ratio backflow/flow $b/a$ necessary for an absolute
instability to happen in unbounded flows (see \cite{HM} and
references therein).

 Eventually, we systematically explore the two regimes of the optimum wavenumber $k_c$
as a function of the amplitude of the backflow $b$. When there is little to no backflow
(small $b$), $k_c$ is nearly constant $k_c\simeq 0.5$ (Fig.~\ref{sm_}
(c), Fig.~\ref{artres} (a)). As $b$ is increased, $k_c$ has sharp
increase. The optimum wavenumber reaches a regime where is grows
continuously with $b$. For smaller values of $d'$ the flow is
stabilised as $b$ is increases, which can lead to no maximum at all
for $\sigma(k)$. This generalises the distinction found in the previous section between intermediate and turbulent area in term of wavelength.

\subsection{Two components base flow $\bar{V}_x,\bar{V}_z$ \label{res2}}

  This case is described by Orr--Sommerfeld--Squire equations (\ref{oss}) and
(\ref{sq}). The same approach as the one-component case is followed here. The main characteristics are presented on a typical case in the first subsection, then a parametric study is conducted.

\subsubsection{General properties}

The case investigated is that of a typical intermediate zone.
The eigenvalue problem is solved for a typical case without backflow ($a=0.7$, $d=3.9$ at $R=300$, $b=0$):
The effect of this parameter on the spanwise sweep is discussed in
the next section. The parameters for $\bar{V}_z$
are found by fitting the DNS results with our model function ($a_z=0.08$
and $d_z=1.8$, see Fig.~\ref{profb1} (c)).

\begin{figure}\centerline{\includegraphics[height=4.2cm]{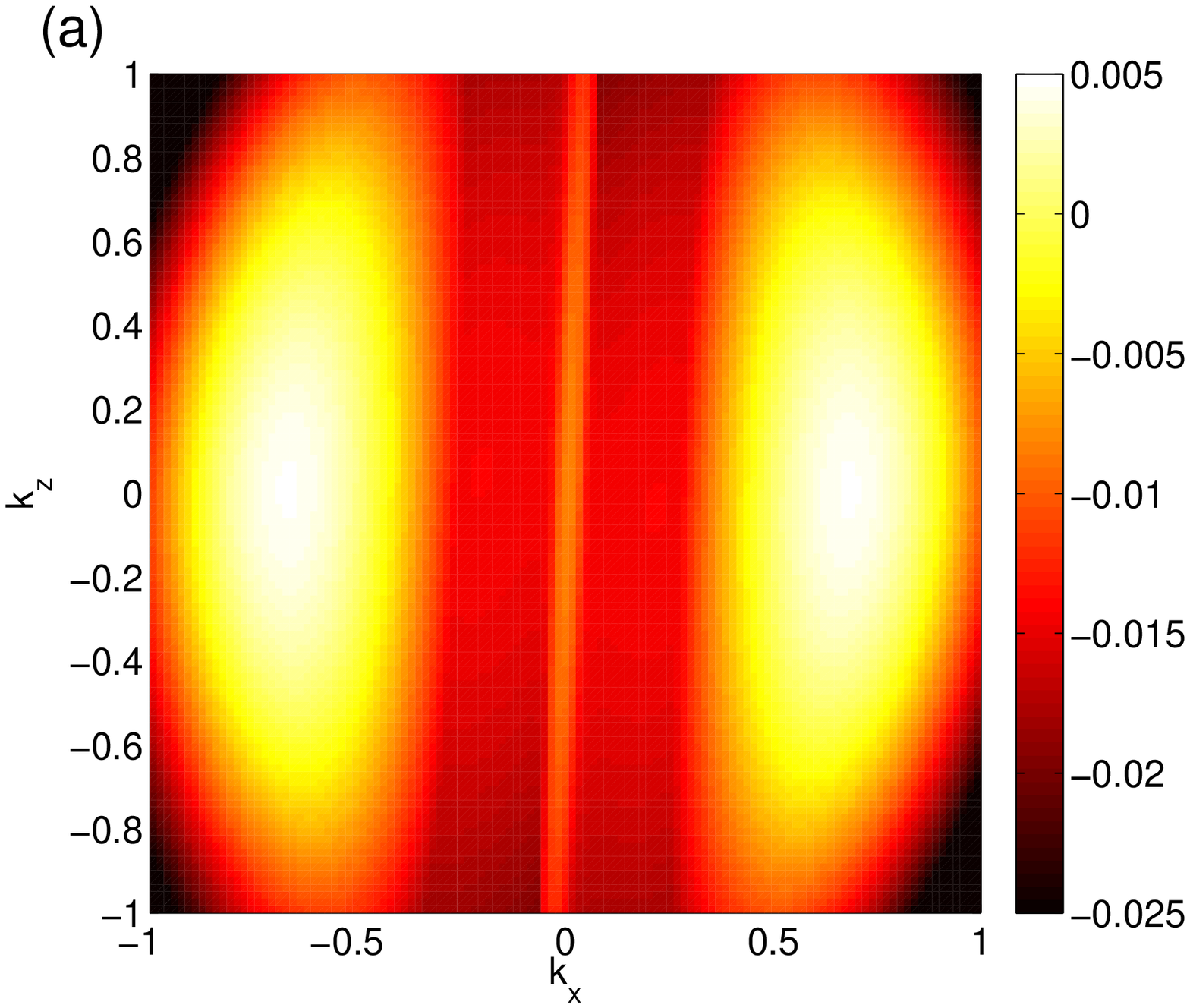}
\includegraphics[height=4.2cm,clip]{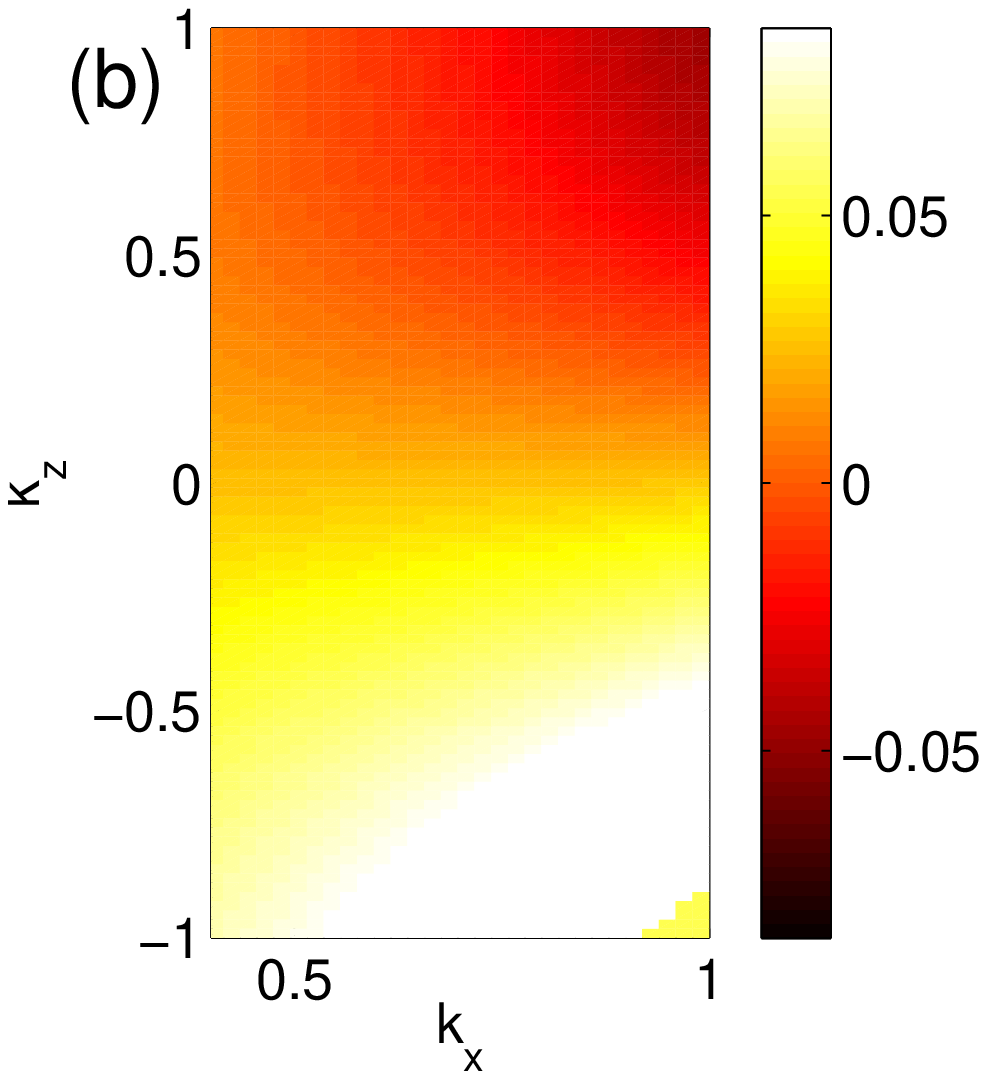}}
\caption{For parameters ($a_x=0.7$, $a_z=0.06$, $s_x=-0.6$, $s_z=-0.1$, $d_x=3.9$, $d_z=1$, $R=300$):
(a) Growth rate $\sigma$ vs wavenumbers $k_x$ $k_z$.
(b): group velocity $c_z$ vs $k_x$ and $k_z$. }\label{figz}
\end{figure}

The growth rate and the group velocities are computed
for these parameters. Their colour levels are displayed as functions of wavenumbers $k_x,k_z$ in
figure~\ref{figz}. The most unstable mode is an eigenmode of the
Orr--Sommerfeld equation. Squire equation is not necessary for the
modal stability analysis of this system. This rules out
three dimensional effects other than advection at the linear order.

  The area for which the background flow is unstable appears in light yellow and white in figure~\ref{figz} (a). Along the $k_x$ axis, the one-component behaviour ($\bar{V}_z=0$) is approximately found (see figure~\ref{artres} (a) for
comparison). The flow is unstable for a given range of wavevectors
$-0.5\lesssim k_z \lesssim 0.5$: it is stabilised by viscosity
at larger wavenumbers $k_z$. The spanwise baseflow, although
unstable by itself does not manage to destabilise the streamwise
baseflow. As detailed in the former section, this is an effect of
the plane Couette baseflow. This can be seen in
figure~\ref{artres} (b).

  The spanwise component of the baseflow $\bar{V}_z$ has a moderate effect on $c_x$:
the behaviour is approximately unchanged when compared to the one
component base flow case (Fig.~\ref{artres} (c)). Its effect is comparable to that of a spanwise sweep in boundary layers. Colour levels
of group velocity $c_z$ are displayed in figure~\ref{figz} (b). It
ranges from $-0.1$ to $0.1$ in that zone of the spectrum, one finds $c_z\sim0.03$ at $\sigma_m$, which compares well to the average of the spanwise baseflow $\int \bar{V}_z\,{\rm d}y=0.032$.

\subsubsection{Parametric dependence}

  We first study the dependence of $\sigma_m(R)$ on $a_z$. Three values are chosen (Fig.~\ref{artres} (b)). For the lowest values of $R$, the maximum growth rate is not changed by much as $a_z$ is increased. The
threshold Reynolds number is unchanged by the spanwise baseflow.

  The systematic dependence of the group velocity $c_z$ on the
spanwise baseflow is studied. The amplitudes $a_z$ and $b_z$
are varied from $0$ to $0.07$ in two typical test cases. The first
one corresponds to $a_x=0.7$, $b_x=0$, there is no streamwise backflow. The second one to $a_x=0.7$ and $b_x=0.35$, there is a streamwise backflow. This tests
the effect of a spanwise baseflow centred in the upper half (the lower half case is symmetrical) of the flow as well as a case with a backflow.

\begin{figure}
\centerline{\includegraphics[height=5cm,clip]{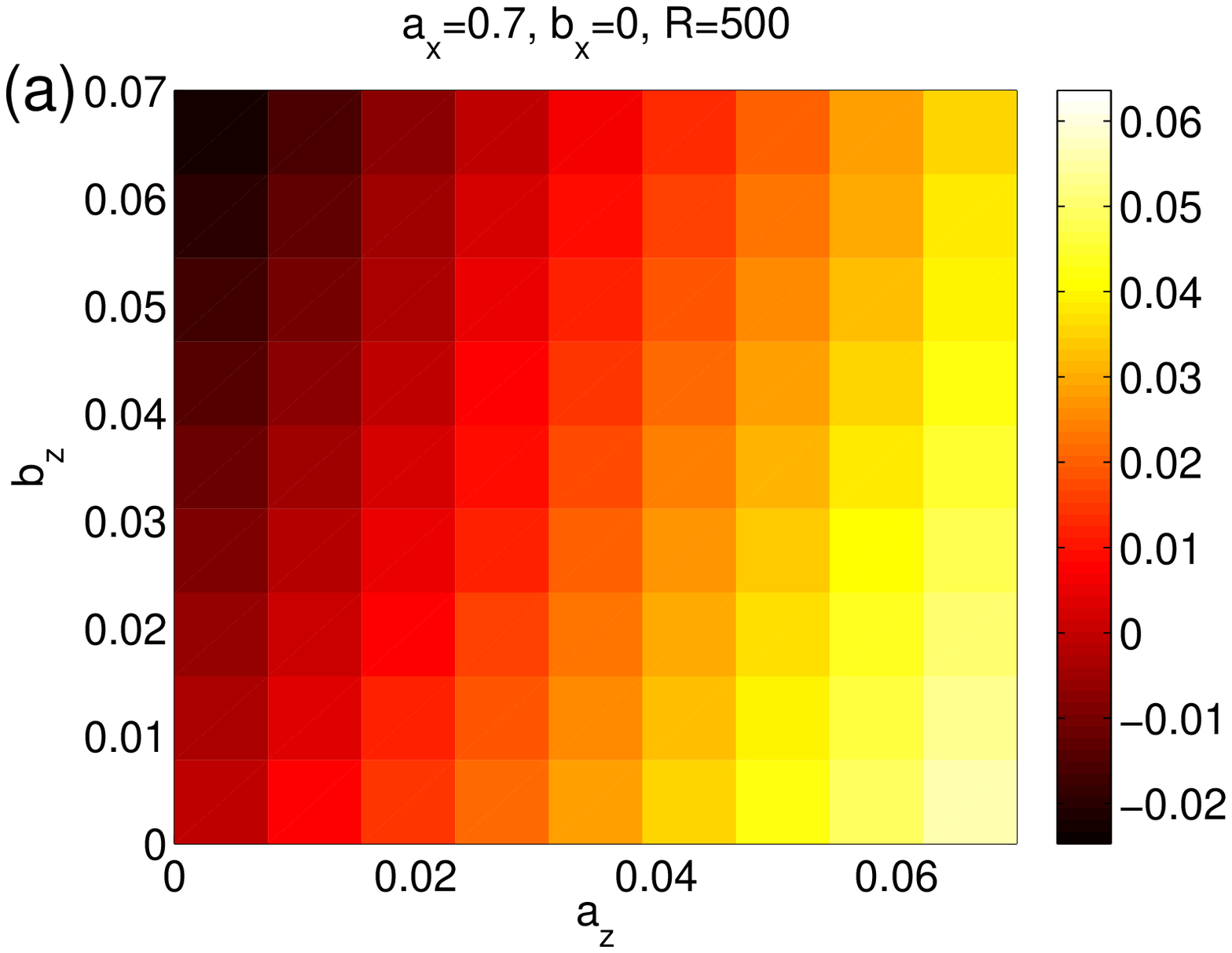}\includegraphics[height=5cm,clip]{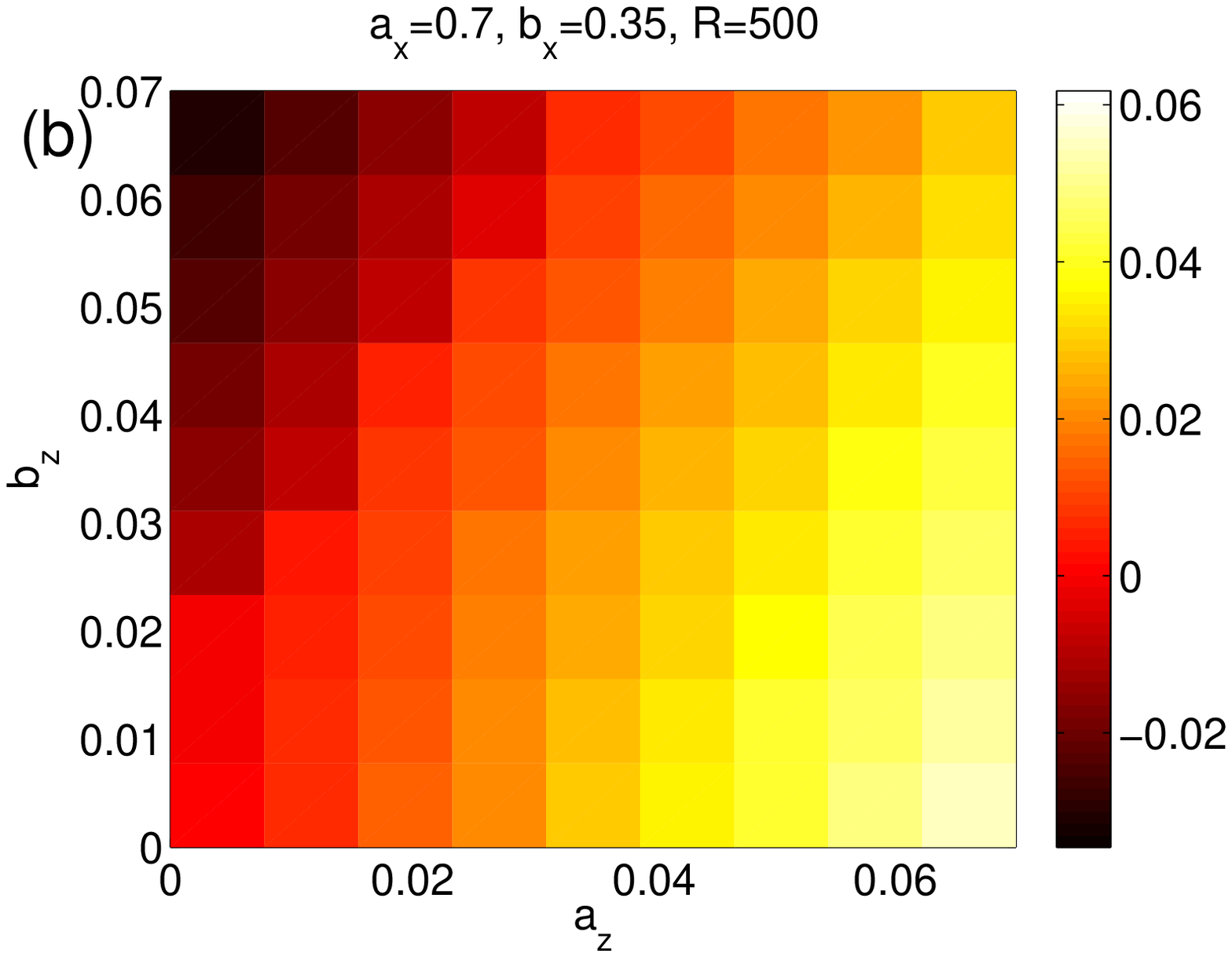}}
\caption{Result of two component stability analysis at $R=300$, where the parameters of $\bar{V}_x$ are fixed and parameters of $\bar{V}_z$ are varied. (a): Spanwise component of the group velocity $c_z(\sigma_m)$ as a function of amplitudes $a_z$ and $b_z$. The inverse shear layer thickness is $d_z=1.8$, the streamwise baseflow has no backflow $a_x=0.7$, $b_x=0$, $d_x=3.9$.
(b): Spanwise component of the group velocity $c_z(\sigma_m)$ as a function of $a_z$ and $b_z$, the inverse shear layer thickness is $d_z=1.8$, the streamwise component of the baseflow has a backflow $a_x=0.7$, $b_x=0.35$, $d_x=3.9$.}
\label{figazbz}
\end{figure}

  The colour levels of the group velocity of the most unstable mode
are displayed in figure~\ref{figazbz}. Like $c_x$, the value of group
velocity $c_z$ is approximately given by the wall normal average of the spanwise baseflow. Indeed, we find $-0.028\le\int_y {\rm d} y
\bar{V}_z\le0.028$. However, one can see an effect of
$\bar{V}_x$: the streamwise background flow is responsible of an dissymmetry of $c_z$ as function of
$a_z$ and $b_z$. If $b_x=0$, the spanwise group velocity $c_z$ takes positive values even for some values of $b_z>a_z$, that is to say, even if the wall normal average of $\bar{V}_z$ is slightly below zero (figure~\ref{figazbz} (a)). If the wall normal average of $\bar{V}_z$ is well below zero, then $c_z$ is negative.
Increasing $b_x$ changes this: $c_z$ can be negative for a large part of the space parameter (figure~\ref{figazbz} (b)). This means that $c_z$ is more sensitive to the values taken by $\bar{V}_z$ for $y>0$ if $\bar{V}_x$ is nonzero in that part of the flow.

\subsection{Global stability analysis \label{glob}}

    We eventually take into account the full streamwise dependance of the velocity streaks in the bands. The problem is considered for qualitative insight on the shape of spectra and modes.
No systematic study is performed in this article.

\subsubsection{Procedure}

 The eigenvalue problem derives from equation~\ref{eqglob}. The eigenmode corresponds to a perturbation of the
streamfunction $\psi$. The streamwise
dependence given by equation~\ref{eqmod2} is used for
the streamwise component of the baseflow with the functional description of equation~\ref{profa1} or~\ref{profa2}. The amplitude and
shear layer thickness are chosen in the range indicated by the DNS
and the local stability analysis: the values $a=0.8$
and $b=0.4$, $d=3.9$, $d'=3$ are used. The wall normal dependence of
equation~\ref{profa1} and~\ref{profa2} is used. the first normalised polynomial $g_0=(\sqrt{315}/16)(1-y^2)^2$
is used for the wall normal dependence of the wall normal component (see \S~\ref{num}).
Its amplitude and streamwise dependence is chosen so as to have
$\partial_x \bar{V}_x+\partial_y \bar{V_y}\simeq0$. This models the
non parallel dependence of the flow. An example of $\bar{V}_x$ in displayed in figure~\ref{glb1} (a).

\subsubsection{Results}

\begin{figure*}
\centerline{\includegraphics[width=17cm,clip]{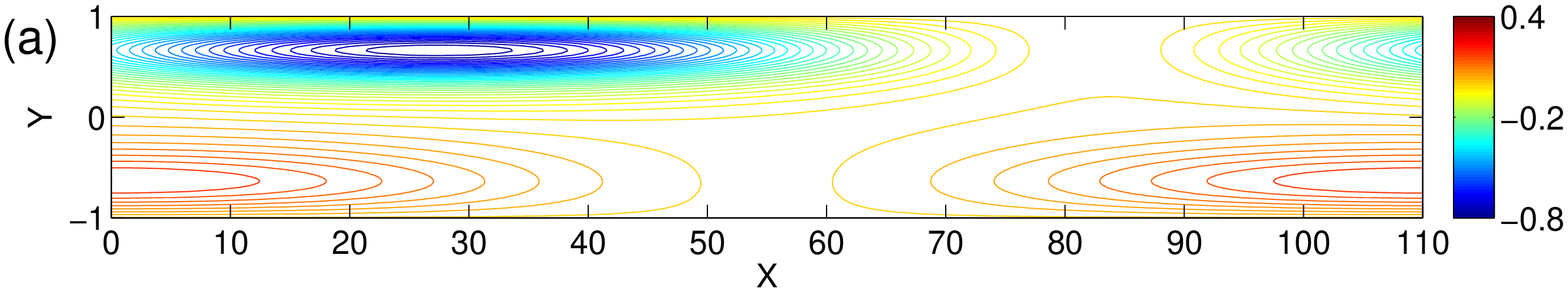}}
\centerline{\textbf{(b)}\hspace{0.1cm}\includegraphics[width=17cm,clip]{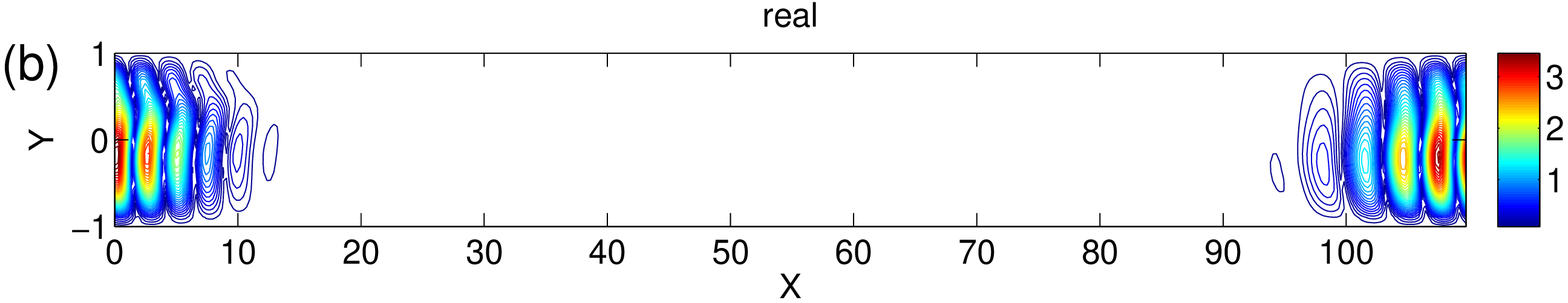}}
\caption{(a) Streamwise baseflow $\bar{V}_x(x,y)$ using parameters $a=0.8$, $b=0.2$, $d=3.9$, $d'=3$. (b): Example of the absolute value of the real part of the most unstable eigenmode $u_y$, calculated at $R=350$, using the base flow of (a).}
\label{glb1}
\end{figure*}

  The analysis is performed for one examples of parameters, at $R=350$, in the middle of the turbulent band existance range.
The typical shape of the eigenmodes can be seen in figure~\ref{glb1}
(b). The shape of the most unstable eigenmode depends very
little on $R$. The real and imaginary part have the same envelope.
Their modulation are in phase quadrature (like sine and a cosine functions). A zero for one part corresponds to an
extremum for the other. The mode is non zero in an area corresponding to the turbulent zone ($95\le x\le 110$, $0\le x \le 15$, remember that the flow is periodical in $x$ figure~\ref{glb1} (b)), that is to say in a zone of the flow where a $v_x<0$ and a $v_x>0$ are superimposed (and shifted by the laminar baseflow). In
the framework of the local stability analysis, this region
corresponds to a zone where the absolute instability is found
(Fig.~\ref{sm_}). In that region, the wavelength of the modulation is comparable to the most unstable wavelength found in the local analysis and the wavelength of perturbations found in the DNS of part $1$ \cite{isp1}.

\begin{figure}
\centerline{\includegraphics[height=6cm,clip]{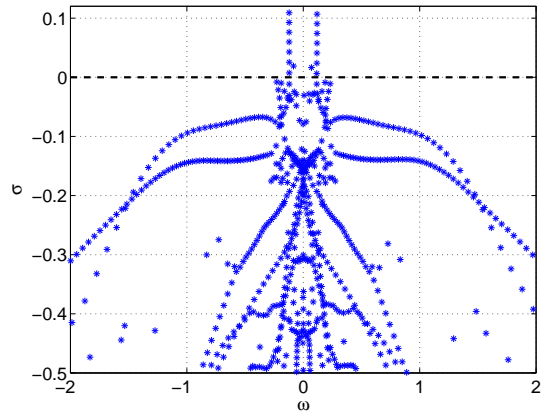}}
\caption{Spectrum (growth rate as a function of the frequency), zoomed on unstable modes. Same regime of parameters as figure~\ref{glb1} $R=350$, $a=0.8$, $b=0.2$, $d=3.9$, $d'=3$, $s=\pm0.6$.}
\label{gbl2}
\end{figure}

  The spectrum for this case is displayed in figure~\ref{gbl2}.
On finds a positive growth rate, while a negative one is expected: the band as a oscillating large scale coherent structure should be globally stable \cite{SL}. The non-trivial structure, the band, should be stable. The source of this discrepancy is unknown. This could result from the necessity of Reynolds stress in global analysis or from the shortcoming of the analysis on a flow in which oscillations are averaged out \cite{SL}.

The frequencies are of order $0.2$, the same order as in the local analysis. Unlike the Hopf bifurcations mentioned here \cite{P,B_w}, where the shedding frequency is systematically measured, a comparison to the DNS to validate this order of magnitude is difficult.

\section{Conclusion \label{disc}}

\subsection{Summary}

This article proposed a linear modeling of the spanwise vorticity formation observed in the velocity streaks of the oblique laminar-turbulent bands of transitional plane Couette flow. Section~\ref{model} contained the preliminary steps of the linear stability analysis: it identified the baseflows on which the analysis would be performed, and how they depend on the streamwise coordinate (\S~\ref{modx}, \S~\ref{param}). In particular it proposed a functional form for the baseflow profile, with a few parameters (amplitude, inverse shear layer thickness, \S~\ref{bf}). It was shown that tuning the parameters would place the analysis in the turbulent zone, the laminar zone, the zones intermediates between laminar and turbulent flow (\S~\ref{param}). The next section provided details on the equations used for linear stability analysis (namely the Orr--Sommerfeld--Squiere system \S~\ref{fw}) as well as on the numerical method used for the stability analysis (\S~\ref{num}).

Section~\ref{res} contains the linear stability analyses. The formalism of local stability analysis (see \cite{HM} for a reference) was first used to analyse two case typical respectively of the inside of the turbulent band and of the zone intermediate between turbulent and laminar zones (\S~\ref{res1}). This section contained the basis of the main finding of the article: the profile found in a velocity streak inside the turbulent zone is absolutely unstable. Meanwhile the one found in a velocity streak inside the intermediate zone is convectively unstable. In that case the group velocity of the mode unstable wavevector is approximately equal to the wall normal average of the baseflow (as found in DNS \cite{isp1}) and the perturbation is pushed toward the turbulent zone. The next two subsections scan the parameter space and generalise this finding: there is a convective to absolute transition as one increased the amplitude of the backflow, that is to say if one travels from the intermediate zone to the turbulent zone. The velocity amplitude and shear layer thickness required for the baseflow to be unstable correspond to the inside of the velocity streak and not to an average profile. This instability occurs in a range of Reynolds number which include that of existence of the bands. Adding a spanwise component to the baseflow only leads to a a spanwise component to the group velocity. Eventually, a global stability analysis is performed which is partially consistent with these findings.

\subsection{The model and the various configurations of Plane Couette flow}

\begin{figure}
\centerline{\includegraphics[width=6.5cm,clip]{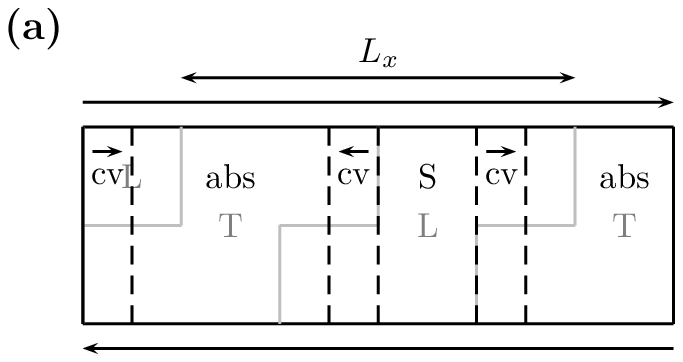}\includegraphics[width=6.5cm]{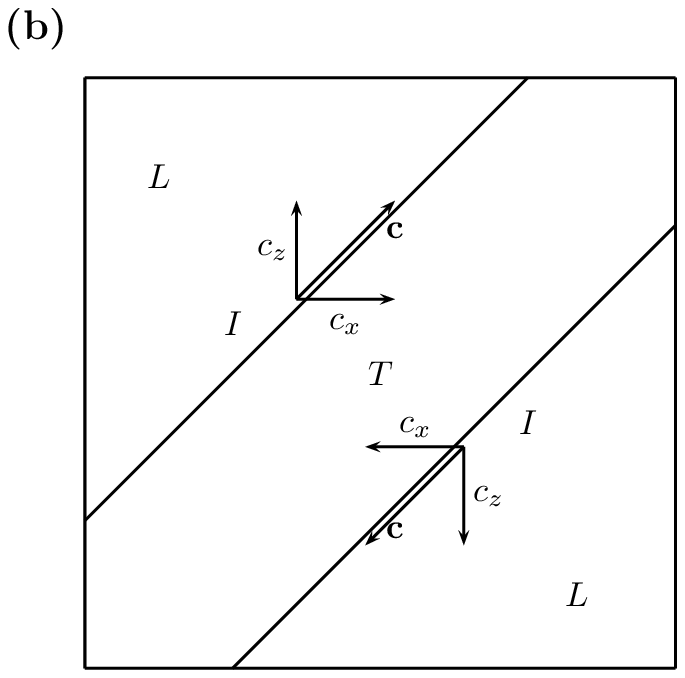}\includegraphics[width=6.5cm]{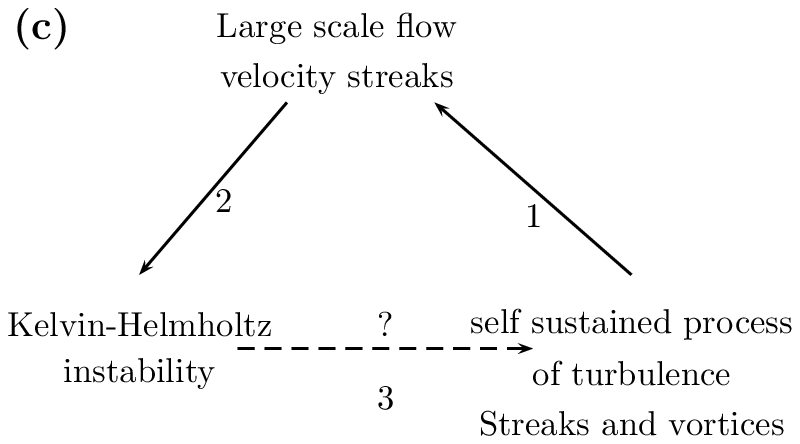}}
\caption{(a): Sketch of the flow in a streamwise wall normal plane indicating the part of the flow occupied by turbulence (denoted by T),
, where the flow is stable (denoted by S), convectively unstable (denoted by cv) and absolutely unstable (denoted by abs),
as well as the direction of the two component group velocity, deduced from linear analysis of section~\ref{res1}.
(b) Sketch of the flow in a streamwise spanwise plane indicating the zones and the directions of streamwise and spanwise group velocities
(c): Possible self-sustaining cycle for the turbulent oblique bands}
\label{cvabs}
\end{figure}

One can use the results on the parametric analysis of of the component baseflow (\S~\ref{res1}) along with information on where these parameters are typically found in the flow (\S~\ref{param}) in order to construct a sketch including the laminar zone, the intermediate zones and the turbulent zone and indicating where the flow is stable, convectively unstable and absolutely unstable (Fig.~\ref{cvabs} (a)). In particular, we use the sampled amplitude of the flow $a$ and backflow $b$ as a function of the streamwise position (Fig.~\ref{profb2} (a)) in regard with the information on whether the flow is stable as a function of flow amplitude $a$ (Fig.~\ref{sm}), and how the amplitude of the backflow $b$ modifies this stability (Fig.~\ref{sm_} (a)) and how $b$ can make the flow absolutely unstable (Fig.~\ref{sm_} (b)). We will only consider one half of the flow ($0\le x\le 55$ Fig.~\ref{profb2} (a)), the other half only corresponds to an inversion of the role of $a$ and $b$. We also only consider reasonable amplitudes of the inverse shear layer thickness ($d,d'\ge 3$). One can see that in the laminar zone $0\le x\le 15$, one finds $a,b\le 0.5$ (Fig.~\ref{profb2} (a)), and the baseflow is stable no matter what (Fig.~\ref{sm}), as reported on the sketch (Fig.~\ref{cvabs} (a)). In the intermediate zone $15\le x \le 40$ (Fig.~\ref{profb2} (a)), $a$ is large enough for the flow to be unstable, while $b\lesssim 0.2$ is small enough so that the flow is convectively unstable, but not absolutely unstable in the range of Reynolds number of existence of the bands $320\lesssim R \lesssim 420$ (Fig.~\ref{sm_} (a,b)). Eventually, in the turbulent zone, $40\le x\le 55$, one can see that $a$ and $b$ are large (Fig.~\ref{profb2} (a)), so that one is fully in the range of parameter where the flow is absolutely unstable (Fig.~\ref{sm} (b)). One can then reconstruct the second part of the flow by inverting the roles of $a$ and $b$, and obtain the full sketch (Fig.~\ref{cvabs} (a)).

Using the results of section~\ref{res2}, where a two component baseflow $(\bar{V}_x,\bar{V}_z)$ was considered, one can also create a sketch of the direction of the two component group velocity $(c_x,c_z)$ depending on the zone of the flow (laminar, intermediate or turbulent). In that section, it was shown that one had $c_z\simeq \int \bar{V}_z\,{\rm d}y$, moreover section~\ref{res1} showed that we also had $c_x\simeq \int \bar{V}_x\,{\rm d}y$. This means that we can draw the same sketch as figure~9 of \cite{isp1}, but in the intermediate and turbulent zone only. This indicates us that that the perturbations arising in the intermediate zone are advected along the band, as was found in direct numerical simulation.

\subsection{Discussion}

  The results on PCF can be discussed in view of the sustaining cycle of metastable puffs proposed by Shimizu \& Kida \cite{SK}. They noted that the turbulence inside the puff creates unstable profiles in the trailing edge. These profiles evolved in a manner similar to a Kelvin-Helmholtz instability of a vorticity sheet to create azimuthal vorticity \cite{SK,DWK}. In the meta-stable puff regime, this vorticity is advected toward the puff \cite{DWK,SK}. They argued that the vorticity reaching the puff excites turbulence and closes the cycle, however, they did not propose a framework in which to place this cycle. The results from direct numerical simulations (part $1$ \cite{isp1}) confirmed that a similar vorticity formation could be found in PCF and pushed the processing further, in particular by measuring the advection velocity as a function of streamwise position. The present study shows that a stability analysis similar to the one performed by Shimizu \& Kida could explain spanwise vorticity formation. Again, it pushed their approach further by performing a local stability analysis for different baseflow, representing different streamwise positions going from the laminar zone to the turbulent zone. By uncovering a convective to absolute transition, the present study showed that an explanation of the type proposed by Shimizu \& Kida and Duguet \emph{et al.} \cite{SK,DWK} could possibly be rephrased in the well known quantitative framework of self-oscillating modes. It is very likely that a study such as this one could be transposed to pipe flow.

  In the case of PCF, an hypothetical cycle is proposed in figure~\ref{cvabs} (c). Three stages are proposed. The description starts from the self sustaining process (SSP) of turbulence has been studied and described in DNS, models \cite{W} and experiments \cite{DD}. The streaks give the unstable profiles while the streaks streamwise vortices yield, after averaging, the large scale flow around the spots and bands (arrow 1). This article and the phenomenological studies  focus on the second arrow of the cycle \cite{isp1}. The velocity streaks are unstable and spanwise vorticity is created. The large scale flow is responsible for the advection of that vorticity along the band. The absolute/convective transition of this instability as one enters the coherent structure of the band draws a parallel between the sustainment of the weakly turbulent band and the sustainment of self-oscillating structures and their transition to turbulence \cite{HM}. It is very likely that the classical streaks instability undergoes such a convective/absolute transition as well \cite{W}, which brings more weight to this type of scenario. However, this part of the cycle (third arrow) is more qualitative and remains to be quantitatively explained. Approaches such as the study of wave/mean flow interaction, common in geophysical fluid dynamics, could bring some insight on this stage of the cycle \cite{valis}. Indeed, they aim at explaining and computing the value of Reynolds stress felt by the flow at large scale based on the rather irregular waves created and advected at small scale. This calls for detailed non-linear analysis of the interaction of these two instability modes in the ``swept'' context of plain Couette flow. The case of plane Couette flow perturbed by a wire, which is not swept, is probably closer to fed back self-oscillating structures \cite{DD}.

  The stability analysis showed that this mechanism functioned in a range of Reynolds number which largely included that of existence of the turbulent bands. This means that it can contribute to the sustainment of the bands. It is more than likely that the mechanisms for laminar hole formation are solely responsible for the appearance of laminar holes as $R$ is decreased below $R_{\rm t}$. Meanwhile, a competition between a sustainment mechanism of the band such as the one described here and hole formation at low Reynolds numbers could be used to explain that bands maintain themselves up to $R_{\rm g}$. This could be done in a manner similar to what has been done to explain spot expansion or retraction \cite{DLS}. This could also be done from another point of view, \emph{i.e.}, that of simplified models (chaotic or stochastic) of such flows \cite{BPPF,B}. This behaviour is generally termed Spatiotemporal intermittency and results from the competition between local transient chaos and contamination from neighbours regenerating this chaos, as it is the case here. Then, knowing the direction and celerity of advection tells which symmetry and which advection should be favoured. The identification of the creation of vorticity shows another source of noise (or local chaos), besides of the streaks/vortices interaction usually involved. These results could be used to improve existing models. In particular, the fact that this sustainment mechanism involves the direction of advection of perturbations by a large scale flow could be used to explain the sustainment of turbulence in an oblique shape.

  These scenarios, together with possible explanation on the distance between two bands and description of the failure of turbulence\cite{m,M11} move toward a better understanding of the intriguing oblique band regime.

\section*{acknowledgments}
The author thanks Y. Duguet and P. Huerre for helpful discussions. He also acknowledges the hospitality of the INLN in Universit\'e Nice Sophia Antipolis where part of the research and manuscript where done. The comments from the anonymous reviewers helped improve the presentation. The author also acknowledge the GRADE language service of Goethe Univerist\"at, Frankfurt, for helpful comments and corrections on the manuscript.


\begin{thebibliography}{00}
\bibitem{prigent}A. Prigent, G. Gr\'egoire, H. Chat\'e, O. Dauchot, \emph{Long-Wavelength modulation of turbulent shear flows}, Physica D \textbf{174} 100--113 (2002).
\bibitem{RM}J. Rolland, P. Manneville, \emph{Ginzburg--Landau description of laminar-turbulent oblique band formation in transitional plane Couette flow}, E. Phys. J. B \textbf{80}, 529--544 (2011).
\bibitem{BT07}D. Barkley, L. Tuckerman, \emph{Mean flow of turbulent-laminar patterns in plane Couette flow}, J. Fluid Mech. \textbf{576} 109--137 (2007).
\bibitem{BT11}L. Tuckerman, D. Barkley, \emph{Patterns and dynamics in transitional plane Couette flow}, Phys. Fluids \textbf{23} 041301 (2011).
\bibitem{lJ}A. Lundbladh A.V. Johansson, \emph{Direct simulation of turbulent spots in plane Couette flow}, J. Fluid Mech. \textbf{229} 499--516 (1991).
\bibitem{BPPF}D. Barkley, \emph{Simplifying the complexity of pipe flow}, Phys. Rev. E \textbf{84},
016309 (2011).
\bibitem{B}D. Barkley, \emph{Modeling the transition to turbulence in shear flows},
J. Phys. Conf. Series, \textbf{318}, 032001 (2011).
\bibitem{SK}M. Shimizu, S. Kida, \emph{A driving mechanism of a turbulent puff in Pipe flow.}, Fluid Dyn. Res. \textbf{41}, 045501 (2009).
\bibitem{DWK}Y. Duguet, A. Willis, R. Kerswell, \emph{Slug genesis in cylindrical pipe flow}, J. Fluid Mech. \textbf{663}, 180--208  (2010).
\bibitem{isp1}J. Rolland, \emph{Formation of spanwise vorticity in oblique turbulent bands of transitional plane Couette flow, part 1: numerical experiments}, Eur. J. Mech. B Flu. \textbf{50}, 52--59 (2015).
\bibitem{DLS}Y. Duguet, O. Le ma\^itre, P. Schlatter, \emph{
Stochastic and deterministic motion of a laminar-turbulent interface in a shear flow},
 Phys. Rev. E \textbf{84}, 066315 (2011).
\bibitem{JP} J. Philip, P. Manneville, \emph{
From temporal to spatiotemporal dynamics in transitional  plane Couette flow},Phys. Rev E \textbf{83} 036308 (2011).
\bibitem{shi}L. Shi, M. Avila, B. Hof, \emph{
Scale invariance at the onset of turbulence in Couette flow}, Phys. Rev. Let. \textbf{110}, 204502 (2013).
\bibitem{ispspot}J. Rolland, \emph{Turbulent spots growth in plane Couette flow: statistical study and secondary instability}, Fluid Dyn. Res. \textbf{46}, 015512 (2014).
\bibitem{ATK} H. Adia, T. Tsukahara, Y. Kawaguchi, \emph{Development of a turbulent spot into a stripe pattern in plane Poiseuille flow}, (2011).
\bibitem{V}E. Villermaux, \emph{On the role of viscosity in shear instabilities}, Phys. Fluids \textbf{10} 368--373 (1998).
\bibitem{BS}R. Betchov, A. Szewczyk, \emph{Stability of a shear layer between parallel streams}, Phys. Fluids \textbf{6} 1391--1396, (1963).
\bibitem{W}F. Waleffe, \emph{On a self sustaining process in shear flows}, Phys. Fluids \textbf{9}, 883--900 (1996).
\bibitem{KJUP}G. Kawahara, J. Jimenez, M. Uhlmann, A. Pinelli, \emph{
Linear instability of a corrugated vortex sheet - a model for streak instabity},
J. Fluid Mech. \textbf{483}, pp. 315--342 (2002).
\bibitem{schhu}W. Schoppa, F. Hussain, \emph{Coherent structure generation in near-wall turbulence},
J. Fluid Mech. \textbf{453}, 57--108 (2002).
\bibitem{PRL}Y. Duguet, P. Schlatter, \emph{
Oblique Laminar-turbulent interfaces in plane shear flows}
Phys. Rev. Let. \textbf{110}, 034502 (2013).
\bibitem{lag}M. Holzner, B. Song, M. Avila, B. Hof, \emph{Lagrangian Approach to laminar-turbulent interfaces in transitional pipe flow}
J. Fluid Mech. \textbf{723}, 140--162 (2013).
\bibitem{DR} P.J. Schmid, D.S. Henningson, \emph{Stability and transition in shear flows}, Springer (2001).
\bibitem{HM}P. Huerre, \emph{Open shear flows instabilities}, in Perspectives in fluid dynamics, 159--229, G.K. Batchelor, H.K. Moffatt , M.G. Worster eds. (2000).
    \bibitem{metal}M. Marquillie U. Erhenstein, J.P. Laval, \emph{Instability of streaks in wall turbulence with adverse pressure gradient}
J. Fluid Mech. \textbf{681}, pp. 205--240 (2011).
    \bibitem{B_w}D. Barkley, \emph{
Linear analysis of the cylinder wake mean flow}, Europhys. Lett. \textbf{75}, 5, 750--756 (2006).
\bibitem{P}B. Pier, \emph{On the frequency selection of finite-amplitude vortex shedding in the cylinder wake},
J. Fluid Mech. \textbf{458}, 407--417 (2002).
\bibitem{SL}D. Sipp, A. Lebedev, \emph{Global stability of base and mean flows: a general approach and its applications to cylinder and cavity flows}
J. Fluid Mech. \textbf{593}, 333-358 (2007).
    \bibitem{cole66} D. Coles, \emph{Transition in circular Couette flow}, J. Fluid Mech. \textbf{21}, 385--425
(1965), Ch. Van Atta, \emph{Exploratory measurements in spiral turbulence}, J. Fluid Mech. \textbf{25}, 495--512 (1966).
    \bibitem{gibs}J. Gibson, J. Halcrow, P. Cvitanovi\'c, \emph{
Visualizing the geometry of state space in plane Couette flow},
J. Fluid Mech. \textbf{611}, 107--130 (2008).
\bibitem{MR} P. Manneville, J. Rolland, \emph{On modelling transitional turbulent flows using under-resolved direct numerical simulations: the case of plane Couette flow},
Theor. Comput Fluid Dyn. \textbf{25}, 407--420, (2010).
\bibitem{dsc10}Y. Duguet, P. Schlatter, D.S. Henningson, \emph{Formation of turbulent patterns near the onset of transition in plane Couette flow},
J. Fluid Mech. \textbf{650}, 119--129 (2010).
\bibitem{BDT} L.S. Tuckerman, D. Barkley, O. Dauchot, \emph{Instability of uniform turbulent plane Couette flow: spectra, probability distribution functions and $K$--$\Omega$ closure model}, in Seventh IUTAM Symposium on Laminar-Turbulent Transition, edited by P. Schlatter and D. Henningson, \textbf{18}, 59--66 (Springer, New York, 2010).
\bibitem{lm}M. Lagha, P. Manneville, \emph{Modeling transitional plane Couette flow},
Eur. Phys. J. B \textbf{58} 433--447 (2007).
\bibitem{conv}R.S. Schechter and D.M. Himmelblau, \emph{
Local potential and system stability}, Phys. Fluids \textbf{8}, 1431--1437 (1965).
\bibitem{chqz}C. Canuto, M. Hussaini, A. Quateroni, T. Zang, \emph{Spectral Methods : Fundamentals in single domains}, Springer (2006).
\bibitem{linalg}G. H. Golub, C. F. Van Loan, \emph{Matrix computations}, Johns Hopkins University Press (1996).
\bibitem{phd}J. Rolland, \emph{\'Etude num\'erique \`a petite et grande
\'echelle de la bande laminaire-turbulente
de l'\'ecoulement de Couette plan
transitionnel}, phd thesis, (2012).
\bibitem{DD}O. Dauchot, F. Daviaud, \emph{Streamwise vortices in plane Couette flow}
Phys. Fluids \textbf{7}, 901--903 (1995).
\bibitem{ACC}C. Arratia, C.P. Caulfield, J.M. Chomaz, \emph{Transient perturbation growth in time dependent mixing layers}, J. Fluid Mech. \textbf{717}, 90--133 (2013).
\bibitem{cb}C. Cossu, L. Brandt, \emph{On tollmien--Schlichting-like waves in streaky boundary layers},
Eur. J. Mech. B flu. \textbf{23}, 815--833 (2004).
\bibitem{m} P. Manneville \emph{Spatiotemporal perspective on the decay of turbulence in wall-bounded flows}, Phys. Rev. E \textbf{79} 025301 (2009).
\bibitem{M11}P. Manneville, \emph{On the decay of turbulence in plane Couette flow}, Fluid Dyn. Res. \textbf{43}, 065501 (2011).
\bibitem{BT99}D. Barkley, L.S. Tuckerman, \emph{
Stability analysis of perturbed plane Couette Flow},
Phys. Fluids \textbf{11}, 1187--1195 (1999).
\bibitem{valis}G.K. Valis, \emph{Atmospheric and oceanic fluid dynamcs}, Cambridge university press (2006).
\end{thebibliography}
\end{document}